\documentclass[a4paper,12pt]{article}
\usepackage{amsfonts}
\usepackage{latexsym}
\usepackage{amsmath}
\usepackage{amssymb}
\usepackage{amssymb}
\usepackage{slashed}
\usepackage{upgreek}
\usepackage{mathrsfs}
\usepackage{bbm}

\usepackage[toc,page]{appendix}

\usepackage{amsthm}

\theoremstyle{remark}

\usepackage[utf8]{inputenc}
\usepackage[english]{babel}

\theoremstyle{definition}

\newcommand*{\defeq}{\mathrel{\vcenter{\baselineskip0.5ex \lineskiplimit0pt
                     \hbox{\scriptsize.}\hbox{\scriptsize.}}}%
                     =}

\allowdisplaybreaks

\usepackage{relsize}
\usepackage{graphicx}
\usepackage{comment}

\renewcommand{\thefootnote}{\fnsymbol{footnote}}

\def\appendix#1{\addtocounter{section}{1}\setcounter{equation}{0}
\renewcommand{\thesection}{\Alph{section}}
\section*{Appendix \thesection\protect\indent \parbox[t]{11.15cm}{#1}}
\addcontentsline{toc}{section}{Appendix \thesection\ \ \ #1}}

\font\mybb=msbm10 at 11pt

\def\bb#1{\hbox{\mybb#1}}

\def\bR {\bb{R}}

\def\bC {\bb{C}}

\def\gX{\Gamma\mkern-4.0mu X}

\def\gY{\Gamma\mkern-2.0mu Y}

\def\gQ{\Gamma\mkern-4.0mu Q}

\def\sX{\slashed {X}}
\def\sgX{\slashed {\gX}}
\def\sY{\slashed {Y}}
\def\sgY{\slashed {\gY}}

\def\sgQ{\slashed {\gQ}}

\newcommand{\bea}{\begin{eqnarray}}
\newcommand{\eea}{\end{eqnarray}}

\usepackage[usenames,dvipsnames,svgnames,table]{xcolor}	
\usepackage[normalem]{ulem}				
\usepackage{microtype}
\usepackage[colorlinks, citecolor=blue, linkcolor=blue, urlcolor=Maroon, filecolor=Maroon, linktocpage=true]{hyperref} 

\usepackage{chngcntr}
\counterwithout{equation}{section}

\begin{document}

\begin{center}
\vspace*{-1.0cm}
\begin{flushright}
\end{flushright}


\vspace{2.0cm} {\Large \bf TCFHs, hidden symmetries and M-theory backgrounds } \\[.2cm]

\vskip 2cm
G.\,  Papadopoulos and  E.\, P\'erez-Bola\~nos
\\
\vskip .6cm


\begin{small}
\textit{Department of Mathematics
\\
King's College London
\\
Strand
\\
 London WC2R 2LS, UK}
 \\*[.3cm]
 \texttt{george.papadopoulos@kcl.ac.uk}
 \\
\texttt{edgar.perez$\underline{~}$bolanos@kcl.ac.uk}
\end{small}
\\*[.6cm]

\end{center}

\vskip 2.5 cm

\begin{abstract}
\noindent
We present the TCFH of 11-dimensional supergravity and so demonstrate that the form bilinears of supersymmetric solutions satisfy a generalisation of the conformal Killing-Yano equation with resepct to the TCFH connection. We also compute the Killing-St\"ackel, Killing-Yano and closed conformal Killing-Yano tensors of all spherically symmetric  M-branes that include the M2-brane, M5-brane, KK-monopole and pp-wave and demonstrate that their geodesic flows are completely integrable by giving all independent conserved charges in involution.
We then find that all form bilinears of pp-wave and KK-monopole solutions  generate (hidden) symmetries  for spinning particle probes propagating on these backgrounds. Moreover, there are Killing spinors such that some of the 1-, 2- and 3-form bilinears of the M2-brane solution also generate symmetries for spinning particle probes. We also explore the question on whether the form bilinears are sufficient to prove the integrability of particle probe dynamics on 11-dimensional supersymmetric backgrounds.
\end{abstract}



\newpage

\renewcommand{\thefootnote}{\arabic{footnote}}

\section{Introduction}

 Killing-St\"ackel (KS) tensors and  (conformal) Killing-Yano  ((C)KY) forms have a long and distinguished history in general relativity as they have been used to investigate the integrability and separability properties of many
classical  equations, like the geodesic, Hamilton-Jacobi, Klein-Gordon, Dirac and Maxwell equations, on black hole spacetimes, see selected references  \cite{carter-b}-\cite{lun} and reviews \cite{revky, frolov}.  In particular KS tensors generate  (hidden) symmetries for relativistic particle probes  propagating on gravitational backgrounds  and so symmetries of the geodesic flow. While KY forms, which can be thought of as the ``square root'' of KS tensors,    generate (hidden) symmetries for spinning particle probes \cite{bvh} propagating on gravitational backgrounds \cite{gibbons}.

More recently, it has been demonstrated in \cite{gptcfh} that the conditions imposed by the Killing spinor equations (KSEs) on the (Killing spinor) form bilinears of any supergravity theory, which may include higher order curvature corrections, can be arranged as a twisted covariant form hierarchy (TCFH) \cite{jggp}.  This means that these conditions can be written as
\bea
{\cal D}_X^{\cal F}\Omega= i_X {\cal P}+ X\wedge {\cal Q}~,
\label{tcfh}
\eea
for every spacetime vector field $X$, where $\Omega$ is a multiform with components the form bilinears,  ${\cal P}$ and ${\cal Q}$  are  appropriate multi-forms which depend on the bilinears and the fields of the theory.  Note that  $X$ also denotes  the associated 1-form constructed from the vector field $X$ after using  the spacetime metric $g$, $X(Y)=g(X,Y)$. Furthermore ${\cal D}^{\cal F}$ is a connection on the space of forms which depends on the fluxes ${\cal F}$ of the supergravity theory that it is not necessarily form degree preserving. A consequence of the TCFH is that the form bilinears $\Omega$ satisfy a generalisation of the  CKY equation with respect to  ${\cal D}^{\cal F}$ connection as one can easily verify by skew-symmetrising and taking the contraction with respect to the spacetime metric  of (\ref{tcfh}). This raises the question on whether the form bilinears generate symmetries for appropriate probes propagating on supersymmetric backgrounds. This question has been explored before in 4- and 5-dimensional minimal supergravities \cite{ebgp} as well in type II 10-dimensional supergravities \cite{lggpjp}.

One  purpose of this paper is to present the full TCFH of 11-dimensional supergravity. We shall find that the reduced holonomy of the minimal\footnote{See \cite{gptcfh} for the definition of these connections.} TCFH connection is included in $SO(10,1)\times GL(517)\times GL(495)$ while the reduced holonomy of the maximal TCFH connection is included in  $GL(528)\times GL(496)$. The latter holonomy is the same as that of the maximal connection of IIA and IIB TCFH \cite{lggpjp}. Then we shall explore the question on whether the TCFH conditions can be identified with the invariance conditions
of a probe action under transformations generated by the form bilinears. As the supersymmetric backgrounds of 11-dimensional supergravity have not been classified, we shall focus our investigation on the M-brane solutions\footnote{These have been instrumental in the understanding of string dualities \cite{hulltown, town}.} which include the M2- and M5-branes as well as the pp-wave and KK-monopole.

Before we proceed with the investigation of the TCFH for M-branes, we shall give the KS tensors and KY forms associated with the complete integrability of the geodesic flow of spherically symmetric M-brane solutions, i.e. those that depend on a harmonic function with one centre. The geodesic equations of these backgrounds are separable in angular variables. Here we shall present all independent conserved charges which are in involution. Moreover we shall demonstrate that a relativistic particle probe  propagating on spherically symmetric M-branes admits an infinite number of hidden symmetries generated by KS tensors. In addition, we shall find that the spinning particle probe action  admits $2^8$, $2^7$ and $2^4$ symmetries generated by KY forms on the pp-wave, M2-brane and M5-brane backgrounds, respectively.  Spinning particle probes exhibit enhanced worldline supersymmetry propagating on the KK-monopole.

  After this, we shall return to investigate under which conditions the form bilinears of M-brane backgrounds, which may now depend of a general harmonic function and so they are not necessarily spherically symmetric, generate symmetries for spinning particle type of probes.  For this we match the  conditions required for a transformation generated by the form bilinears to leave a spinning particle probe action invariant with the TCFH conditions on the form bilinears. We shall find that all form bilinears of pp-wave and KK-monopole backgrounds generate symmetries for the spinning particle probes.  This is because as a consequence of the TCFH and the vanishing of the 4-form field strength for these solutions,  the form bilinears are covariantly constant with respect to the Levi-Civita connection. Furthermore we demonstrate that there are Killing spinors such that the 1-form, 2-form and 3-form bilinears of the M2-brane are KY forms and so generate symmetries
 for spinning particle probes propagating on this background.  A similar analysis for the M5-brane reveals that only the 1-form bilinear generates symmetries
 for spinning particle probes. To demonstrate  these results, we have computed all the form bilinears of M-brane backgrounds using spinorial geometry  \cite{ggp}.

This paper is organised as follows. In section 2, we present the TCFH of 11-dimensional supergravity and give the reduced holonomy of TCFH connections.  In section 3, we give the KS and KY tensors of spherically symmetric M-brane backgrounds and prove the complete integrability of their geodesic flows. In section 4, we identify the form bilinears of M-branes that generate symmetries for probe actions, and in section 5 we give our conclusions. In appendix A, we give the form bilinears of the M5-brane. In appendix B, we explore the symmetries of spinning particle probes with 4-form couplings.

\section{The TCFH of D=11 supergravity}

 The
 supercovariant connection  of $11-$dimensional supergravity \cite{julia} is
\bea
D_\mu=\nabla_\mu+{1\over 288}(\Gamma_\mu{}^{\nu_1\nu_2\nu_3\nu_4}F_{\nu_1\nu_2\nu_3\nu_4}-8F_{\mu\nu_1\nu_2\nu_3}\Gamma^{\nu_1\nu_2\nu_3})~,
\eea
where $\nabla$ is the spin connection of the spacetime metric, $F$ is the 4-form field strength of the theory and $\epsilon$ is a $\mathfrak{spin}(10,1)$ Majorana spinor.  The reduced holonomy of supercovariant connection on generic backgrounds is included in $SL(32, \bR)$ \cite{hull, duff, gpdt}.

Supersymmetric backgrounds with $N$ Killing spinors, $\epsilon^r$, $r=1, \dots, N$, are those that   admit $N$ linearly independent  solutions to the KSE, $D_\mu\epsilon^r=0$. Given $N$ Killing spinors, one can construct the form bilinears
\bea
&&
f^{rs}=\langle \epsilon^r, \epsilon^s \rangle~,~~~
k^{rs}_\mu=\langle \epsilon^r, \Gamma_\mu\epsilon^s\rangle~,~~~
\omega^{rs}_{\mu\nu}=\langle \epsilon^r, \Gamma_{\mu\nu}\epsilon^s\rangle~,~~~
\varphi^{rs}_{\mu_1\mu_2\mu_3}=\langle \epsilon^r, \Gamma_{\mu_1\mu_2\mu_3}\epsilon^s\rangle ~,~~~
\cr
&&
\theta^{rs}_{\mu_1\mu_2\mu_3\mu_4}=\langle \epsilon^r, \Gamma_{\mu_1\mu_2\mu_3\mu_4}\epsilon^s\rangle~,~~~
\tau^{rs}_{\mu_1\mu_2\mu_3\mu_4\mu_5}=\langle \epsilon^r, \Gamma_{\mu_1\mu_2\mu_3\mu_4\mu_5}\epsilon^s\rangle~.~~~
\label{11bi}
\eea
Note that the form bilinears $k$, $\omega$ and $\tau$ are symmetric in the exchange of $\epsilon^r$ and $\epsilon^s$ while the rest are skew-symmetric. There is no a classification of supersymmetric solutions of 11-dimensional supergravity. However there are many partial results.  For example the maximally supersymmetric solutions have been classified in \cite{jffgp} and the KSE has been solved for one Killing spinor in \cite{jgsp1, jgsp2, ggp}, see review \cite{ggprev} for the current state of the art.

The TCFH of $11-$dimensional supergravity for the form bilinears which are symmetric in the exchange of the two Killing spinors has been given in \cite{gptcfh}.  Here we shall present the TCFH for all form bilinears.
The TCFH  of $11-$dimensional supergravity expressed in terms of the minimal connection ${\cal D}_\mu^{\cal F}$ reads
\bea
\label{11dtcfh}
&&
{\cal D}_\mu^{\cal F}k_\nu\defeq \nabla_\mu k_\nu={1\over6}F_{\mu\nu\alpha\beta}\omega^{\alpha\beta}-{1\over 6!}{}^\ast F_{\mu\nu\rho_1\rho_2\rho_3\rho_4\rho_5}\tau^{\rho_1\rho_2\rho_3\rho_4\rho_5}~,~~~
\cr
&&
{\cal D}_\mu^{\cal F}\omega_{\nu_1\nu_2}\defeq \nabla_\mu \omega_{\nu_1\nu_2}-{1\over 2\cdot 3!}F_{\mu\rho_1\rho_2\rho_3}\tau^{\rho_1\rho_2\rho_3}{}_{\nu_1\nu_2}=-{1\over 3}F_{\mu\nu_1\nu_2\rho}k^{\rho}
\cr
&&
\qquad\qquad-{1\over 2 \cdot 3!}\tau_{[\mu\nu_1}{}^{\rho_1\rho_2\rho_3}F_{\nu_2]\rho_1\rho_2\rho_3}+{1\over 3 \cdot 4!}g_{\mu[\nu_1}\tau_{\nu_2]}{}^{\rho_1\rho_2\rho_3\rho_4}F_{\rho_1\rho_2\rho_3\rho_4}~,~~~
\cr
&&
{\cal D}_\mu^{\cal F} \tau_{\nu_1\nu_2\nu_3\nu_4\nu_5} \defeq \nabla_\mu \tau_{\nu_1\nu_2\nu_3\nu_4\nu_5}+5F_{\mu[\nu_1\nu_2\nu_3}\omega_{\nu_4\nu_5]}-{5\over 6}{}^\ast F_{\mu[\nu_1\nu_2\nu_3|\rho_1\rho_2\rho_3|}\tau_{\nu_4\nu_5]}{}^{\rho_1\rho_2\rho_3}=
\cr
&&
\qquad\qquad-{1\over 6}{}^\ast F_{\mu\nu_1\nu_2\nu_3\nu_4\nu_5\rho}k^{\rho}+{5\over 2}F_{[\mu\nu_1\nu_2\nu_3}\omega_{\nu_4\nu_5]}-{5\over 6}\tau_{[\mu\nu_1}{}^{\rho_1\rho_2\rho_3}{}^\ast F_{\nu_2\nu_3\nu_4\nu_5]\rho_1\rho_2\rho_3}
\cr
&&
\qquad\qquad-{10\over 3}g_{\mu[\nu_1}\omega^{\rho}{}_{\nu_2}F_{\nu_3\nu_4\nu_5]\rho}-{5\over 18}g_{\mu[\nu_1}\tau_{\nu_2}{}^{\rho_1\rho_2\rho_3\rho_4}{}^\ast F_{\nu_3\nu_4\nu_5]\rho_1\rho_2\rho_3\rho_4} ~,
\cr
&&
{\cal D}_\mu^{\cal F}f\defeq \nabla_\mu f={1\over 18}F_{\mu\nu_1\nu_2\nu_3}\varphi^{\nu_1\nu_2\nu_3}~,~~~
\cr
&&
{\cal D}_\mu^{\cal F}\varphi_{\nu_1\nu_2\nu_3}\defeq \nabla_\mu \varphi_{\nu_1\nu_2\nu_3}-{3\over 4}F_{\mu[\nu_1|\rho_1\rho_2|}\theta^{\rho_1\rho_2}{}_{\nu_2\nu_3]}={1\over 6}g_{\mu[\nu_1}F_{\nu_2|\rho_1\rho_2\rho_3|}\theta^{\rho_1\rho_2\rho_3}{}_{\nu_3]}
\cr
&&
\qquad\qquad-{1\over 36} {}^\ast F_{\mu\nu_1\nu_2\nu_3\rho_1\rho_2\rho_3}\varphi^{\rho_1\rho_2\rho_3}-{1\over 2}F_{[\mu\nu_1|\rho_1\rho_2|}\theta^{\rho_1\rho_2}{}_{\nu_2\nu_3]}-{1\over 3}F_{\mu\nu_1\nu_2\nu_3}f~,~~~
\cr
&&
{\cal D}_\mu^{\cal F} \theta_{\nu_1\nu_2\nu_3\nu_4} \defeq\nabla_\mu \theta_{\nu_1\nu_2\nu_3\nu_4}-{1\over 3}{}^\ast F_{\mu[\nu_1\nu_2\nu_3|\rho_1\rho_2\rho_3|}\theta^{\rho_1\rho_2\rho_3}{}_{\nu_4]}+3F_{\mu[\nu_1\nu_2|\rho|}\varphi^{\rho}{}_{\nu_3\nu_4]}=
\cr
&&
\qquad\qquad{1\over 18}g_{\mu[\nu_1}{}^\ast F_{\nu_2\nu_3\nu_4]\rho_1\rho_2\rho_3\rho_4}\theta^{\rho_1\rho_2\rho_3\rho_4}-{5\over 18}{}^\ast F_{[\mu\nu_1\nu_2\nu_3|\rho_1\rho_2\rho_3|}\theta^{\rho_1\rho_2\rho_3}{}_{\nu_4]}
\cr
&&
\qquad\qquad-g_{\mu[\nu_1}F_{\nu_2\nu_3|\rho_1\rho_2|}\varphi^{\rho_1\rho_2}{}_{\nu_4]}+{5\over 3} F_{[\mu\nu_1\nu_2|\rho|}\varphi^{\rho}{}_{\nu_3\nu_4]}~,~~~
\label{11tcfh}
\eea
where for simplicity we have suppress the indices $r$ and $s$ on the form bilinears with label the independent Killing spinors.
In our conventions $\epsilon_{0123456789\natural}=-1$, ${}^\ast F_{\mu_1\cdots \mu_7}= {1\over4!}\epsilon_{\mu_1\cdots \mu_7}{} ^{\nu_1\cdots \nu_4}F_{\nu_1\cdots \nu_4}$ and $\Gamma_{\natural}\defeq \Gamma_{0\dots 9}$, where $\natural$ denotes the 11th direction. Clearly the equations above are of the form stated in  (\ref{tcfh}), where $\Omega$ is the multiform spanned by the form bilinears (\ref{11bi}), ${\cal Q}$ can be read from the terms in the right hand side of  (\ref{11tcfh}) that explicitly contain the spacetime metric $g$ and ${\cal P}$ is spanned by the remaining terms  in the right hand side of (\ref{11tcfh}). Clearly (\ref{11tcfh}) provides a geometric interpretation  of the conditions induced by the KSE on the form bilinears as it relates them to a generalisation of the CKY equations.

Viewing ${\cal D}_\mu^{\cal F}$ as degree non-preserving connection on k-forms, $k=0,1,2,3,4,5$, the reduced holonomy of ${\cal D}_\mu^{\cal F}$ factorises as the connection preserves
the subspaces of $k$-degree forms for $k=1,2,5$   and  for $k=0, 3, 4$, i.e. it preserves the subspaces   of the form bilinears which are symmetric and skew-symmetric under the exchange of the two Killing spinors.  This is also the case for the maximal  connection defined in \cite{gptcfh} which we do not consider here in detail.  In addition, ${\cal D}_\mu^{\cal F}$ preserves the subspace of 1-forms, and the subspace of 2- and 5-forms, and acts trivially on 0-forms.  As a result the reduced holonomy of  ${\cal D}_\mu^{\cal F}$  is included in
$SO(10,1)\times GL(517)\times GL(495)$ group. Note that the reduced holonomy of the maximal connection is included in  $GL(528)\times GL(496)$ as it does not preserve the subspace of 1-forms but instead it mixes them with the subspace of 2- and 5-forms and it acts non-trivially
on 0-forms. The reduced holonomy of the maximal connection is the same as that of the maximal TCFH connections of type IIA and type IIB supergravities \cite{lggpjp}.  Of course for special backgrounds the holonomy of  ${\cal D}^{\cal F}$  reduces further.

\section{Symmetries  of probes on  M-brane backgrounds}

\subsection{Symmetries and integrability}

We shall begin with a summary of the key properties of Killing-St\"ackel (KS) tensors and Killing-Yano (KY) forms. As this has already appeared in the form required here elsewhere \cite{lggpjp}, we shall be brief. Consider the action of a relativistic particle probe propagating on a spacetime $M$ with metric $g$ 
\bea
A={1\over2}\int\, dt\, g_{\mu\nu}\, \dot x^\mu\, \dot x^\nu~,
\label{gact}
\eea
where $\dot x$ denotes the derivative of the coordinate $x$ with respect to    $t$. The equations of motion are those of the geodesic flow on $M$ with affine parameter $t$. Given a rank $k$  Killing-St\"ackel (KS) tensor on $M$, i.e.  a symmetric $(0,k)$ tensor $d$ on $M$ which satisfies that equation
\bea
 \nabla_{(\mu} d_{\nu_1\nu_2\cdots\nu_k)}= 0~,
 \label{cks}
\eea
where  $\nabla$ is the Levi-Civita connection of  $g$, the action (\ref{gact}) is invariant under the infinitesimal transformations
\bea
\delta x^\mu= \epsilon d^\mu{}_{\nu_1\cdots \nu_{k-1}} \dot x^{\nu_1} \cdots \dot x^{\nu_{k-1}}~,
\label{kstran}
\eea
with parameter $\epsilon$. The associated conserved charge is
\bea
Q(d)= d_{\nu_1\nu_2\cdots\nu_k}\, \dot x^{\nu_1}\, \dot x^{\nu_2}\cdots \dot x^{\nu_k}~.
\label{ccd}
\eea
For $k=1$, $d$ is a Killing  vector field. The symmetrised tensor product of two KS tensors is also a KS tensor. Hidden symmetries are those generated by rank $k\geq 2$ KS tensors $d$  with $d\not=g$.

A conformal Killing-Yano (CKY) tensor is a $k$-form on a spacetime $M$ with metric $g$ which satisfies the condition
\bea
\nabla_\mu \alpha_{\nu_1\nu_2\cdots\nu_k}={1\over k+1} d\alpha_{\mu\nu_1\dots \nu_k}-{k\over n-k+1} g_{\mu[\nu_1} \delta\alpha_{\nu_2\cdots \nu_k]}~.
\label{cky}
\eea
If $\delta\alpha=0$, then $\alpha$ is a KY form while if $d\alpha=0$, $\alpha$ is a closed conformal Killing-Yano (CCKY) form.  It turns out that if $\alpha$ is KY, then the Hodge dual ${}^*\alpha$ is CCKY form.   KY forms are the ``square roots'' of KS tensors.  In particular  if $\alpha$ and $\beta$ are k-KY  forms, then
$\alpha_{(\mu }{}^{\lambda_1\cdots \lambda_{k-1}} \beta_{\nu) \lambda_1\cdots\lambda_{k-1}}$
is a rank 2 KS  tensor.

A spinning particle probe propagating on a spacetime $M$ with metric $g$ is described by the action
\bea
A=-{i\over2} \int\, dt\, d\theta\,\, g_{\mu\nu}\, D x^\mu\, \dot x^\nu~,
\label{sgact}
\eea
where $t$ and $\theta$ are the even and odd coordinates of the worldline superspace, respectively,    $x$ are worldline superfields $x=x(t, \theta)$ and $D^2=i\partial_t$. Spinning particles  are  supersymmetric extensions of relativistic particles.

Given a KY form, $\alpha$, on $M$,  the infinitesimal transformation
\bea
\delta x^\mu=\epsilon\, \alpha^\mu{}_{\nu_1\cdots \nu_{k-1}} Dx^{\nu_1}\cdots Dx^{\nu_{k-1}}~,
\label{svar}
\eea
with parameter $\epsilon$ leaves the spinning particle action (\ref{sgact})
invariant.   The associated conserved charge is
\bea
Q(\alpha)=(k+1) \alpha_{\nu_1\nu_2\cdots \nu_k} \partial_t x^{\nu_1} Dx^{\nu_2}\cdots Dx^{\nu_k}-{i\over k+1} (d\alpha)_{\nu_1\nu_2\cdots \nu_{k+1}} Dx^{\nu_1} Dx^{\nu_2}\cdots Dx^{\nu_{k+1}}~.
\label{ccalpha}
\eea
Observe that $Q(\alpha)$ is preserved, $DQ(\alpha)=0$, subject to the equations of motion of (\ref{sgact}).
Note that if $d\alpha=0$ and so $\alpha$ is covariantly constant (or equivalently parallel) with respect to the Levi-Civita connection, then
\bea
\tilde Q(\alpha)= \alpha_{\nu_1\nu_2\cdots \nu_k} D x^{\nu_1} Dx^{\nu_2}\cdots Dx^{\nu_k}~,
\eea
is also conserved subject to the field equations of (\ref{sgact}), $\partial_t \tilde Q(\alpha)=0$.  There are several generalisations of the KS and CKY tensors, see e.g. \cite{gggpks}-\cite{sat}.

The commutator  algebra of transformations (\ref{svar})
  generated by spacetime forms has been examined in detail in \cite{phgp}.  Given two symmetries (\ref{svar}) generated by the $k$-form $\alpha$ and $\ell$-form $\beta$, the commutator contains two types of terms. One terms depends  on the Nijenhuis tensor of $\alpha$ and $\beta$ and the other term is the transformation 
  \bea
  \delta x^\mu=\epsilon (\alpha \cdot_s \beta)^\mu{}_{\nu\lambda_1\dots \lambda_{k+\ell-4}} \partial_t x^\nu Dx^{\lambda_1} \dots Dx^{\lambda^{k+\ell-4}}~,
  \label{ksky}
  \eea
  generated by the tensor 
  \bea
  (\alpha \cdot_s \beta)_{\mu\nu \lambda_1\dots \lambda_{k+\ell-4}}=\alpha_{\mu \kappa [\lambda_1\dots \lambda_{k-2}} \beta^\kappa{}_{|\nu| \lambda_{k-1}\dots \lambda_{k+\ell-4}}+\alpha_{\nu \kappa [\lambda_1\dots \lambda_{k-2}} \beta^\kappa{}_{|\mu| \lambda_{k-1}\dots \lambda_{k+\ell-4}]}~, 
  \eea
  where $\epsilon$ is an infinitesimal parameter.
  Clearly if $\alpha$ and $\beta$ are rank 2 KY tensors, then $\alpha \cdot_s \beta$ is a KS tensor.
  In the case that both $\alpha$ and $\beta$ are covariantly constant with respect to the Levi-Civita connection, the Nijenhuis tensor vanishes and so the transformation (\ref{ksky}) is a symmetry of the spinning particle action (\ref{sgact}). This will be the case for all symmetries generated by the form bilinears of pp-wave and KK-monopole solutions.


 Consider a dynamical system with $2n$-dimensional phase space $P$. This  is completely integrable, according to Liouville,  provided  that $P$ admits $n$ independent functions (observables) $Q_r$, $r=1,\dots,n$, including the Hamiltonian, in involution.  $Q_r$ are independent provided that the  map $Q: P\rightarrow \bR^n$, where $Q=(Q_1, \dots, Q_n)$,  has  rank $n$. Moreover $Q_r$ are in involution, iff $
\{Q_r, Q_s\}_{\mathrm{PB}}=0$, i.e. Poisson bracket of any two $Q_r$s' vanishes\footnote{Complete integrability is related to the separability of the equations of motion of a dynamical system. In the phase space coordinates $(Q_1, \dots, Q_n, \psi_1, \dots, \psi_n)$ defined by the charges $Q_1, \dots, Q_n$ and the action-angle coordinates $(\psi_1, \dots, \psi_n)$ adapted to the Hamiltonian vector fields, $X_{Q_i}=\partial_{\psi_i}$, the time evolution of the system is at most linear.}.

Returning to the relativistic particle, the conserved charges  (\ref{ccd}) can be written in phase space variables as
\bea
Q(d)= d^{\nu_1\cdots\nu_k} p_{\nu_1} \cdots p_{\nu_k}~,
\label{ccdb}
\eea
where $p_\mu$ is the conjugate momentum of $x^\mu$ and we have raised the indices of $d$ with the spacetime metric $g$.  These clearly commute with the Hamiltonian $H={1\over2} g^{\mu\nu} p_\mu p_\nu$ as they are constants of motion. Furthermore the Poisson bracket algebra of two constants of motion $Q(d_1)$ and $Q(d_2)$ is $\{Q(d_1), Q(d_2)\}_{\mathrm{PB}}= Q([d_1, d_2]_{\mathrm{NS}})$, where
\bea
([d_1, d_2]_{\mathrm{NS}})^{\nu_1\cdots \nu_{k+\ell-1}}=k d_1^{\mu(\nu_1\cdots \nu_{k-1}} \partial_{\mu} d_2^{\nu_k\cdots \nu_{k+\ell-1})}-\ell d_2^{\mu(\nu_1\cdots \nu_{\ell-1}} \partial_{\mu} d_1^{\nu_k\cdots \nu_{k+\ell-1})}~,
\eea
is the Nijenhuis-Schouten bracket of the KS tensors $d_1$ and $d_2$.
Observe that if $d_1$ is a vector, then $[d_1, d_2]_{\mathrm{NS}}={\cal L}_{d_1} d_2$, i.e. the Nijenhuis-Schouten bracket is the Lie derivative of $d_2$ with respect to the vector field $d_1$. Therefore two charges are in involution provided that the Nijenhuis-Schouten bracket of the associated KS tensors vanishes.

In the examples that follow below, the complete integrability of the geodesic flow of the spacetimes considered is due to the large number of isometries that these spacetimes admit. As the Lie algebra of these isometries is not abelian, the associated conserved charges are not in involution. Nevertheless, it is possible to
use these charges to construct new ones associated  with KS tensors which are in involution, see  the example below.

\subsection{Complete integrability of black hole geodesic flow}

Before we proceed to investigate the symmetries of probes on M-theory backgrounds, let us present some examples. The standard example is the integrability  of the geodesic flow of the Kerr black hole. However more suitable for the results that follow are the examples of Schwarzschild and Reissner-Nordstr\"om black holes in four and higher dimensions. The metric of both these solutions in four dimensions can be written as
\bea
g=-A(r) t^2+ A^{-1}(r) dr^2+ r^2 (d\theta^2+\sin^2\theta d\phi^2)~.
\label{srnmetr}
\eea
The associated geodesic equations of the metric above can be explicitly separated in the stated coordinates. However it is instructive to provide a symmetry argument for the complete integrability of the geodesic equations.

The isometry group of the above backgrounds is $\bR\times SO(3)$.  There are two commuting isometries given by $k_0=\partial_t$ and $k_1=\partial_\phi$ which give rise to the conserved charges $K_0=p_t$ and $K_1=p_\phi$. These together with the Hamiltonian $H={1\over2} g^{\mu\nu} p_\mu p_\nu$ give three conserved charges in-involution. Note that
$[K_r, H]_{\mathrm{NS}}={\cal L}_{k_r} g^{\mu\nu} p_\mu p_\nu=0$, $r=0,1$, as $k_r$ are isometries. It remains to find a fourth conserved charge in involution for the complete integrability of the geodesic system. For this consider the Killing vector fields
\bea
k_1=\partial_\phi~,~~~k_2=-\sin\phi \cot \theta \partial_\phi+ \cos\phi \partial_\theta~,~~~k_3=\cos\phi \cot\theta \partial_\phi+\sin\phi \partial_\theta~,
\label{s2isos}
\eea
which generate the $SO(3)$ isometry group and notice that
$[k_a, k_b]=-\epsilon_{ab}{}{}^c k_c$. Then another conserved charge can be constructed utilising the (quadratic) Casimir operator of the Lie algebra of $SO(3)$ which can be used to construct a symmetric tensor that commutes with all the isometries of the background. As the quadratic Casimir is proportional to the identity matrix in the basis chosen for the Lie algebra, the associated symmetric tensor is
\bea
d=\delta^{ab} k_a\otimes k_b={1\over \sin^2\theta} (\partial_\phi)^2+ (\partial_\theta)^2~.
\label{srncas}
\eea
This is a KS tensor because $k_a$ are Killing vectors.
 Thus
\bea
Q(d)={1\over \sin^2\theta} p_\phi^2+ p_\theta^2~,
\eea
is a conserved charge of the geodesic flow of the metric (\ref{srnmetr}).
It turns out that $K_r, H$ and $Q(d)$ are independent and in involution implying that the geodesic equations are completely integrable for any function $A=A(r)$ in (\ref{srnmetr}).

The metric (\ref{srnmetr}) also admits a CCKY 2-form \cite{frolov}.  This is given by
\bea
\beta=r dt\wedge dr~,
\label{bhccky}
\eea
which can be verified after a computation.  The dual
\bea
\alpha={}^*\beta=r^3 \sin\theta d\theta\wedge d\phi~,
\label{srnky}
\eea
is a KY 2-form. As a result it generates a symmetry for the spinning particle action (\ref{sgact}) given by the infinitesimal variation in (\ref{svar}). There are four additional KY 1-forms constructed from the Killing vector fields $k_0, k_1, k_2, k_3$ using the metric.  All of which generate symmetries for the action (\ref{sgact}).
One can also square the KY tensor (\ref{srnky}) to construct a KS tensor. It turns out that this is not independent from (\ref{srncas}).

The analysis we have done can be extended to black holes in higher than four dimensions. Indeed consider the metric
\bea
g=-A(r) dt^2+ A^{-1}(r) dr^2+ r^2 g(S^n)~,
\label{hsrnmetr}
\eea
where $g(S^n)$ is the round metric on $S^n$ with $n\geq 2$. Again the geodesic equation can be separated in angular coordinates and the geodesic flow is completely integrable.
The above metric admits a $\bR\times SO(n+1)$ group of isometries.  Viewing $S^n$ embedded as the hypersurface, $\sum_i (x^i)^2=1$, in $\bR^{n+1}$, the Killing vectors of the spacetime metric $g$ can be written as
\bea
k_0=\partial_t~,~~~k_{ij}=x_i\partial_j-x_j\partial_i~,~~~i<j~,
\eea
where $i,j=1, \dots, n+1$ and $x_i=x^i$.  Note that $k_{ij}$ are tangent to $S^n$ as $(d(x^2-1))(k_{ij})=2 x_k dx^k(k_{ij})=0$. The associated conserved charges are $Q_0=p_t$ and $Q_{ij}=x_i p_j-x_j p_i$, where $p_i$ is the momentum on $S^n$ and so $x^i p_i=0$. These conserved charges are not in involution. However
\bea
Q_0~,~~~D_m={1\over4} \sum_{i,j\geq n+2-m} (Q_{ij})^2~,~~~m=2, \dots, n+1~,
\label{inch}
\eea
are independent conserved charges of the geodesic flow and in involution which together with the Hamiltonian, $H$, of the geodesic motion
imply the complete integrability of the geodesic flow of the metric (\ref{hsrnmetr}).

To explain the choice of $D_m$ charges in (\ref{inch}), note that $D_{n+1}$ is the Hamiltonian of the geodesic flow on $S^n$ and it is constructed using  the quadratic Casimir operator of $\mathfrak{so}(n+1)$. The $\mathfrak{so}(n+1)$ algebra admits a decomposition
\bea
\mathfrak{so}(2)\subset \mathfrak{so}(3)\subset\cdots\subset \mathfrak{so}(n)\subset \mathfrak{so}(n+1)~.
\eea
The $D_m$ conserved charge is constructed using the quadratic Casimir operator of the $\mathfrak{so}(m)$ subalgebra of $\mathfrak{so}(n+1)$.  At each stage  as the quadratic Casimir operator of $\mathfrak{so}(m)$ is invariant under $\mathfrak{so}(m)$, it is also invariant under the $\mathfrak{so}(m-1)$ subalgebra of $\mathfrak{so}(m)$. Therefore the quadratic Casimir operator of $\mathfrak{so}(m-1)$ commutes with that of $\mathfrak{so}(m)$. As a consequence $D_{m-1}$ is in involution with $D_m$.  This method of constructing observables in involution has been generalised and used in \cite{thimm} to investigate the integrability of geodesic flows on homogeneous manifolds.

Moreover a direct computation reveals that $\beta$ in (\ref{bhccky}) is a CCKY form for the metric (\ref{hsrnmetr}) and therefore its dual $\alpha$ is a KY $n$-form. It turns out that $\beta$ in (\ref{bhccky}) is a CCKY with respect to a metric as in (\ref{hsrnmetr}) with $g(S^n)$ now replaced with the metric, $g(N)$, of any $n$-dimensional manifold $N$ provided it is independent from the coordinates $r$ and $t$.

\subsection{Hidden symmetries and spherically symmetric M-branes}

Next let us turn to investigate the symmetries of relativistic and spinning particle probes described by the actions  (\ref{gact}) and (\ref{sgact}), respectively,  propagating on M-branes.  The focus will be on those  KS and KY
tensors which give rise to conserved charges related to the integrability of the geodesic flow on some of these backgrounds.

\subsubsection{M-theory pp-waves}

The M-theory pp-wave solution is
\bea
g=2 du(dv+ {1\over2} h(y, v) du)+ \delta_{ij} dy^i dy^j~,
\label{ppwave}
\eea
with $F=0$, where $(u,v,y)$ are the coordinates of 11-dimensional spacetime and $h$ is a Harmonic function on $\bR^9$, $\partial_y^2 h=0$. As the $\partial_y^2 h=0$ condition appears in other M-brane backgrounds below,  the solutions of this equation that we shall be considering on $\bR^n$, $n>2$, are
 \bea
 h=q_0+\sum^\ell_{m=1} {q_m\over |y-y_m|^{n-2}}~,~~~q_0=0,1~,
 \label{harm}
 \eea
 where $q_m$ are constants, $|\cdot|$ is the Euclidean norm on $\bR^n$ and $y_m$ are the centres or positions of the harmonic function $h$.

Here we shall investigate the symmetries of probes propagating on a spherically symmetric  pp-wave, i.e. a pp-wave that depends on a harmonic function with one centre. After a coordinate transformation to put the centre at $0$,   $h={q\over |y|^7}$, where $q$ is constant denoting the momentum of the pp-wave.
This solution has an $\bR^2\times SO(9)$ symmetry generated by the Killing vector fields $k_+=\partial_u$, $k_-=\partial_v$ and
 \bea
k_{ij}=y_i \partial_j-y_j\partial_i~,~~~i<j~,
\label{rotvf}
\eea
where $y_i=y^i$. The latter vector fields are generated by
the action of $SO(9)$ on the $y$ coordinates.

Clearly all the above vector fields generate symmetries for the probe action (\ref{gact})  with conserved charges
\bea
Q_\pm=Q(k_\pm)=p_\pm~,~~~Q_{ij}=Q(k_{ij})=y_i p_j-y_j p_i~.
\eea
In addition, one can demonstrate with a direct calculation that
 \bea
 d_{i_1\dotsi_k}= y^{j_1}\dots y^{j_q} a_{j_1\dots j_q, i_1\dots i_k}~,
 \eea
 are KS tensors of the pp-wave spacetime provided that the constant tensor $a$ satisfies the condition\footnote{ For $q=k$, the $(0,2q)$ tensors that lie in the irreducible representation of $GL(9)$ associated with the 2 rows and $q$ columns Young tableau solve the condition on $a$. A similar statement is true for the KS tensors of the M2- and M5-branes below.}
 \bea
 a_{(j_1\dots j_q, i_1)\dots i_k}=a_{j_1\dots (j_q, i_1\dots i_k)}=0~.
 \eea 
 These in turn generate transformations as those in (\ref{kstran}) which leave the action  (\ref{gact}) invariant.  The associated conserved charges are given in (\ref{ccd}) or equivalently in (\ref{ccdb}). It is evident from the above analysis that a probe described by the action (\ref{gact}) and propagating on this pp-wave spacetime, and so the geodesic flow,  admits infinite number of hidden symmetries.  Note that KS and KY tensors on 4-dimensional pp-wave spacetimes have been investigated before, see e.g. \cite{ppks, ppky}.

 Although the probe  (\ref{gact}) admits an infinite number of symmetries propagating on a pp-wave background, it does not immediately imply that the dynamics is completely integrable.  Clearly the conserved charges $Q_{ij}=Q(k_{ij})$ and $Q_\pm=Q(k_\pm)$ generated by the vector fields $k_{ij}$ and $k_\pm$ are not in involution-the Poisson bracket algebra of $Q_{ij}$ is $\mathfrak{so}(9)$.  However $Q_{ij}$ can be used to construct conserved charges which are in involution. In particular, one can show that the 10 conserved charges
\bea
Q_\pm~,~~~D_m={1\over4} \sum_{i,j\geq 10-m} (Q_{ij})^2~,~~~m=2, \dots, 9~,
\eea
are in involution. These  together with the Hamiltonian of the geodesic system $H={1\over2} g^{\mu\nu} p_\mu p_\nu$ give 11 independent conserved charges in involution leading to the complete integrability of the geodesic flow. As in the black hole analysis, $D_9$ is the Hamiltonian of the geodesic flow on $S^8$ which is constructed from the quadratic Casimir operator of $\mathfrak{so}(9)$.

Turning to the investigation of the symmetries of the probe (\ref{sgact}) propagating on a pp-wave, one has to determine the KY tensors of the background. One can verify after some calculation that
\bea
\beta(\varphi)=  y_i  dy^i \wedge\varphi\wedge du\wedge dv~,
\eea
are CCKY forms
for any constant k-form $\varphi$ on $\bR^9$, where $y_i=y^i$. As a result $\alpha(\varphi)={}^*\beta (\varphi)$ are KY forms.  These generate the transformations
(\ref{svar}) which leave the spinning particle action (\ref{sgact}) invariant  with associated conserved charges given in (\ref{ccalpha}). Therefore the probe (\ref{sgact}) propagating on a pp-wave background admits $2^8$ linearly independent conserved charges\footnote{The maximal number of independent KY $k$-forms \cite{kastor} on a $n$-dimensional spacetime is $(n+1)!/((k+1)! (n-k)!)$.} generated by the KY forms $\alpha(\varphi)$.

\subsubsection{M2-branes}

The M2-brane solution \cite{ksmd} can be expressed as
\bea
g=h^{-{2\over3}} \eta_{ab} d\sigma^a d\sigma^b+ h^{{1\over3}} \delta_{ij} dy^i dy^j~,~~~F= \pm d\sigma^0\wedge d\sigma^1\wedge d\sigma^2\wedge d h^{-1}~,
\eea
where $\sigma^a$, $a=0,1,2$, are the worldvolume coordinates of the brane,  $y^i$, $i=1,\dots,8$, are the transverse coordinates and  $h$ is a harmonic function $\partial_y^2 h=0$ on the transverse space $\bR^8$. An explicit expression for $h$ is  as in (\ref{harm}) with $q_0=1$ and $n=8$.

 For the spherically symmetric M2-brane solution that we shall consider in this section   $h=1+{q\over |y|^6}$.  This solution is invariant under the action of the $SO(1,2)\ltimes \bR^3\times SO(8)$ group, where the Poincar\'e group acts on the worldvolume coordinates of the M2-brane while $SO(8)$ acts on the transverse coordinates with standard rotations.  The Killing vector fields are $k_a=\partial_a$, $k_{ab}=\sigma_a\partial_b-\sigma_b \partial_a$ and $k_{ij}=y_i \partial_j-y_j \partial_i$, where $\sigma_a= \eta_{ab} \sigma^b$ and $y^i=y_i$.  It is clear that the probe (\ref{gact}) propagating on this background admits symmetries generated by these vector fields and the associated conserved charges are
\bea
Q_a=Q(k_a)=p_a~,~~~Q_{ab}=\sigma_a p_b-\sigma_b p_a~,~~~Q_{ij}=Q(k_{ij})=y_i p_j-y_j p_i~.
\label{m2ch}
\eea
As for the pp-wave, the probe (\ref{gact}) admits additional symmetries generated by KS tensors.  To find these tensors we use an ansatz which preserves the worldvolume Poincar\'e symmetry of the solution.  Then after some computation one can verify that
\bea
d_{a_1\dots a_{2m}i_1\dots i_k}=  h^{{1\over3} (k-2m)}   y^{j_1}\dots y^{j_q} a_{j_1\dots j_q, i_1\dots i_k} \eta_{(a_1a_2}\dots \eta_{a_{2m-1} a_{2m})}~,
\eea
are KS tensors provided that the constant tensors $a$ satisfy
\bea
a_{(j_1\dots j_q, i_1)\dots i_k}=a_{j_1\dots (j_q, i_1\dots i_k)}=0~.
\eea
These in turn give additional conserved charges (\ref{ccd}) for the relativistic particle probe (\ref{gact}).  Therefore the probe (\ref{gact}), and so the geodesic flow on this M2-brane, admits an infinite number of hidden conserved charges.

The dynamics of the relativistic particle (\ref{gact}) propagating on this M2-brane background, and so the geodesic flow, is completely integrable. Indeed one can verify after some calculation that the conserved  charges
\bea
Q_a~,~~~D_m={1\over4} \sum_{i,j\geq 9-m} (Q_{ij})^2~,~~~m=2, \dots, 8~,
\eea
are in involution. These together with the Hamiltonian of the relativistic particle (\ref{gact})  yield 11 independent conserved charges  in involution.

Next let us turn to investigate the symmetries of the spinning particle probe (\ref{sgact}) propagating on the spherically symmetric M2-brane. Clearly the Killing vector fields  of the M2-brane generate symmetries for the probe (\ref{sgact}). Additional symmetries are generated by the KY forms of this M2-brane. To find these, we adapt an ansatz which is invariant under the worldvolume Poincar\'e  group of  the M2-brane.  Then after some computation, one finds that
\bea
\beta(\varphi)=h^{{1\over6}(k-4)}  y_i dy^i\wedge \varphi\wedge d{\mathrm{vol}}(\bR^{2,1})~,
\eea
are CCKY tensors of the M2-brane for any constant k-form $\varphi$ on $\bR^8$, where  $d{\mathrm{vol}}(\bR^{2,1})$ is the volume form of $\bR^{2,1}$.  As a result
$\alpha(\varphi)={}^*\beta(\varphi)$ are KY tensor and so spinning particle action (\ref{sgact}) is invariant the under transformation (\ref{svar}) generated by $\alpha(\varphi)$.  The associated constants of motion are given in (\ref{ccalpha}).  These KY tensors generate $2^7$ linearly independent hidden symmetries for the action (\ref{sgact}).
.

\subsubsection{M5-branes}

The M5-brane solution \cite{gueven} is
\bea
g=h^{-{1\over3}} \eta_{ab} d\sigma^a d\sigma^b+ h^{{2\over3}} \delta_{ij} dy^i dy^j~,~~~F=\pm  \star_5 dh~,
\label{m5}
\eea
where $\sigma^a$, $a=0,\dots, 5$, are the worldvolume coordinates, $y^i$, $i=1,\dots,5$, are the transverse coordinates, the Hodge duality operation has been taken with respect to the flat metric on the transverse space $\bR^5$ and $h$ is a harmonic function, $\partial_y^2 h=0$, on $\bR^5$.  $h$ is given in (\ref{harm}) with $n=5$ and $q_0=1$.

For the spherically symmetric M5-brane solution that we  consider here, $h$ has one centre and so it can be arranged such that $h=1+{q\over |y|^3}$. Such a solution  admits a $SO(1,5)\ltimes\bR^6\times SO(5)$ isometry group. The Killing vector fields are $k_a=\partial_a$, $k_{ab}= \sigma_a \partial_b-\sigma_b \partial_a$ and $k_{ij}=y_i \partial_j-y_j \partial_i$, where $y_i=y^i$ and $\sigma_a=\eta_{ab} \sigma^b$. The transformations generated by these vector fields leave invariant the relativistic particle action (\ref{gact}) and the associated conserved charges are
\bea
Q_a=p_a~,~~Q_{ab}=\sigma_a p_b-\sigma_b p_a~,~~~Q_{ij}=y_i p_j-y_j p_i~.
\label{m5ch}
\eea
As for M2-branes, relativistic particles  propagating on the above M5-brane background admit additional symmetries associated with KS tensors.
Adapting again an ansatz which is invariant under the  worldvolume Poincar\'e symmetry and after some computation one finds that
\bea
d_{a_1\dots a_{2m}i_1\dots i_k}=  h^{{1\over3} (2k-m)}   y^{j_1}\dots y^{j_q} a_{j_1\dots j_q, i_1\dots i_k} \eta_{(a_1a_2}\dots \eta_{a_{2m-1} a_{2m})}~,
\eea
are KS tensors
provided that the constant tensors $a$ satisfy
\bea
a_{(j_1\dots j_q, i_1)\dots i_k}=a_{j_1\dots (j_q, i_1\dots i_k)}=0~.
\eea
Clearly, these generate infinite many hidden symmetries for the relativistic particle action (\ref{gact}). So  the geodesic flow on the spherically symmetric M5-brane has infinite many conserved charges.

Furthermore, one can show that the dynamics of relativistic particles propagating on this M5-brane is completely integral. Indeed one can verify that the 10 conserved charges
\bea
Q_a~,~~~~D_m={1\over4} \sum_{i,j\geq 6-m} (Q_{ij})^2~,~~~m\geq 2,\dots, 5~,
\eea
are in involution.  These together with the Hamiltonian of (\ref{gact}) yield 11 independent conserved charges in involution as required for complete integrability.

As for the M2-brane, the spinning particle action (\ref{sgact}) admits, in addition to the symmetries generated by the Killing vectors field of the M5-brane, hidden symmetries generated by KY forms. To find these we adapt and ansatz which is invariant under the worldvolume Poincar\'e group of the M5-brane. Then after some computation, one can verify that
\bea
\beta(\varphi)=h^{{1\over3}(k-1)}  y_i dy^i\wedge  \varphi\wedge d{\mathrm{vol}}(\bR^{5,1})~,
\eea
are CCKY forms for any constant k-form $\varphi$ on $\bR^5$. As a result $\alpha(\varphi)={}^*\beta(\varphi)$ are KY forms and so generate symmetries (\ref{svar}) for the spinning particle probe
(\ref{sgact}) with conserved charges (\ref{ccalpha}).  These KY forms generate $2^4$ linearly independent hidden symmetries.

\subsubsection{KK-monopoles}

The KK-monopole solution is
\bea
g=\eta_{ab} d\sigma^a d\sigma^b+g_{(4)}~,~~~g_{(4)}=h^{-1} (d\rho+\omega)^2+ h \delta_{ij} dy^i dy^j~,
\label{kk}
\eea
with $F=0$, where $\sigma^a$, $a=0,\dots, 6$, are the worldvolume coordinates and  $g_{(4)}$ is in general the Gibbons-Hawking hyper-K\"ahler metric with $\star_3 dh=d\omega$.  $h$ is a harmonic action on $\bR^3$, $\partial_y^2 h=0$. An expression for $h$ can be found in (\ref{harm}) for $n=3$.

 Here we shall consider  the KK monopole solution  with $g_{(4)}$ the Taub-NUT metric. In such a case   $h$ has one centre and so one can set without loss of generality $h=1+{q\over |y|}$. The isometry group of the solution is $SO(1,6)\ltimes \bR^7\times SO(2)\times SO(3)$. As for the solutions investigated already, the Killing vector fields generated by the Poincar\'e subgroup acting on the worldvolume coordinates are $k_a=\partial_a$ and $k_{ab}= \sigma_a \partial_b-\sigma_b \partial_a$.  To give the vector fields generated by the $SO(2)\times SO(3)$ subgroup, write the  Taub-NUT metric $g_{(4)}$ is angular coordinates as
\bea
g_{(4)}=h^{-1} (d\rho+ q \cos\theta d\phi)^2+ h  \big (dr^2+r^2 (d\theta^2+\sin^2\theta d\phi^2)\big)~,
\eea
with $|y|=r$. Then the Killing vector fields generated by $SO(2)\times SO(3)$  are given by
\bea
&&\tilde k_0=\partial_\rho~,~~~\tilde k_1=\partial_\phi~,~~~\tilde k_2=-\sin\phi \cot \theta \partial_\phi+ \cos\phi \partial_\theta+ q {\sin\phi\over \sin\theta}\partial_\rho~,~~~
\cr
&&\tilde k_3=\cos\phi \cot\theta \partial_\phi+\sin\phi \partial_\theta-q{\cos\phi\over\sin\theta}\partial_\rho~.
\label{ghisos}
\eea
The $SO(3)$ Killing vector fields are as in (\ref{s2isos}) with the addition of a component along $\partial_\rho$ because $\omega$ is not invariant under (\ref{s2isos}) but instead it is invariant up to a gauge transformation.

As the relativistic particle action (\ref{gact}) is invariant under  all these isometries, the associated conserved charges are $Q_a=p_a$, $Q_{ab}=\sigma_a p_b-\sigma_b p_a$ $\tilde Q_0=p_\rho$ and $\tilde Q_r= \tilde k_r^i p_i$,  $r=1,2,3$,  where $\sigma_a=\eta_{ab} \sigma^b$ and $\tilde k_r$ are given in (\ref{ghisos}).  The background admits several KS tensors. As the solution is a product $\bR^{6,1}\times N$, where $N$ is the Taub-NUT manifold, one can consider
the KS tensors of  $\bR^{6,1}$ and $N$ separately. One can easily verify that the symmetric tensors
\bea
d_{a_1\dots a_{k}}=  \sigma ^{b_1}\dots \sigma^{b_q} c_{b_1\dots b_q, a_1\dots a_k} ~,
\eea
are KS tensors\footnote{There is a systematic investigation of KS tensors on Minkowski spacetime as well as on some black hole spacetimes. For example there is a 20 dimensional space of rank 2 conformal KS  tensors on 4-dimensional Minkowski spacetime, see for a summary \cite{anderson}. But the approach adopted here  suffices.} provided that the constants $c$ satisfy $c_{b_1\dots (b_q, a_1\dots a_k)}=0$. $N$ also admits three KS tensors given in \cite{ggpr} which we shall not explicitly state them here. They are constructed from the K\"ahler forms and the KY tensor of $N$ given below. All these isometries and KS tensors generate infinite number of symmetries for the relativistic particle action (\ref{gact}).

The dynamics of the relativistic particle probe
(\ref{gact}), or equivalently the geodesic flow, is completely integrable on this background.  Indeed
the commuting isometries of the KK-monopole are $k_a={\partial\over \partial \sigma^a}$, $\tilde k_0=\partial_\rho$ and $\tilde k_1=\partial_\phi$.  These together with the Hamiltonian of the geodesic system give ten conserved charges in involution. There is an additional independent conserved charge in involution associated  to the quadratic Casimir of $SO(3)$ and constructed using the Killing vector fields (\ref{ghisos}) as
\bea
D={1\over \sin^2\theta} (p_\phi-q \cos\theta p_\sigma)^2+p_\theta^2+ q^2 p_\sigma^2~,
\eea
 which proves the statement. The integrability of the geodesic flow on the Taub-NUT space has been known for sometime, see \cite{ggpr}.

 The $SO(1,6)\ltimes \bR^7\times SO(2)\times SO(3)$ isometries  mentioned above  also generate symmetries for the spinning particle probe (\ref{sgact}) propagating on the KK-monopole background. Such probes have additional hidden symmetries. For example, it is well known that $g_{\mathrm{GH}}$  is a hyper-K\"ahler metric for any (multi-centred)  harmonic function $h$. The associated K\"ahler forms are
\bea
\kappa_{(i)}= (d\rho+\omega)\wedge dy^i-{1\over2} h\,\epsilon^i{}_{jk} dy^j\wedge dy^k~.
\eea
These 2-forms are anti-self-dual on the transverse directions of the KK-monopole, parallel with respect to the Levi-Civita connection  and the associated complex structures satisfy the algebra of imaginary unit quaternions. As a result  these K\"ahler forms can be thought as  KY tensors and so generate symmetries (\ref{svar})
for the probe action (\ref{sgact}) with conserved charges (\ref{ccalpha}).

The KK monopole  admits additional KY tensors.  These are those of $\bR^{6,1}$ and those of $N$. Observe that
\bea
\alpha={1\over k!} \big(\chi_{ a_1\dots a_k}+ \sigma ^{b}\varphi_{b, a_1\dots a_k}\big)\, d\sigma^{a_1}\wedge \dots\wedge d\sigma^{a_k}~,
\eea
are KY tensors of  $\bR^{6,1}$ for any constant tensors $\chi, \varphi$ with the latter to satisfy $\varphi_{b, a_1\dots a_k}=\varphi_{ [b, a_1\dots a_k]}$ \cite{kastor}. If
$N$ is the Taub-Nut space, it is known \cite{ggpr}, see also \cite{visinescu},  that
\bea
 \tilde\alpha= (d\rho+ q \cos \theta d\phi)\wedge dr+ r (2r+q) (1+{r\over q}) \sin\theta d\theta\wedge d\phi~,
\eea
is the KY form.  All these  KY forms generate symmetries for the spinning probe action  (\ref{sgact}). Incidentally the three KS tensors mentioned above are constructed from squaring $\tilde\alpha$ with $\kappa_{(i)}$.

\section{Hidden symmetries from the TCFH}

\subsection{Hidden symmetries and M-theory pp-waves}
Assuming that the pp-wave propagates in the 5th direction\footnote{This choice of worldvolume directions for the pp-wave, and those of the rest of M-branes below, may seem unconventional. But they are convenient as they are aligned with the basis used for the description of spinors in the context of spinorial geometry that we utilise to solve the conditions on the Killing spinors.} and allowing the pp-wave metric (\ref{ppwave}) to depend on a (multi-centred) harmonic function as in (\ref{harm}) with $q_0=0$ and $n=9$, the Killing spinors of the background are constant, $\epsilon=\epsilon_0$ and satisfy the condition\footnote{All gamma matrices considered from  in section 4 and appendix A are in a frame basis.}
$\Gamma_{05}\epsilon_0=\pm \epsilon_0$.  To solve this condition, we shall use spinorial geometry and write $\epsilon_0=\eta+e_5\wedge \lambda$, where $\eta$ and $\lambda$ are Majorana\footnote{Note that the reality condition on $\epsilon$ in the spinorial geometry basis is $\Gamma_{6789}* \epsilon=\epsilon$ which in turn implies that $\eta$ and $\lambda$ are real as well.} $\mathfrak{spin}(9)$ spinors, i.e. $\eta, \lambda\in \Lambda^*(\bR\langle e_1, \dots, e_4\rangle)$ with the reality condition imposed by the anti-linear operation $\Gamma_{6789}*$. Choosing the plus sign in the condition for $\epsilon_0$, this can be solved to yield
$\epsilon=\epsilon_0=\eta$, i.e.  $\Gamma_{05}\epsilon_0= \epsilon_0$ implies that $\lambda=0$.

Given the solution of the condition on $\epsilon$ implied by the KSE, it is straightforward to compute all the bilinears of the background. In particular one finds that $f^{rs}=0$ for all Killing spinors and the rest of the form bilinears (\ref{11bi}) can be written as
\bea
(e^0-e^5)\wedge \phi^{rs}
\eea
where
\bea
\phi^{rs}={1\over k!}\, \langle\eta^r,\Gamma_{i_1\dots i_k}\eta^s\rangle_H \, e^{i_1}\wedge\dots \wedge e^{i_k}~,~~~k=0,1,2,3,4~,
\eea
  $\langle\cdot ,\cdot\rangle_H$ is the Hermitian inner product  restricted on the Majorana representation of $\mathfrak{spin}(9)$ and $i_1, \dots, i_k=1, 2, 3, 4, 6, 7, 8, 9, \natural$. Moreover $(e^0, e^5, e^i)$ is a pseudo-orthonormal frame such that $-e^0+e^5=\sqrt2\, du$, $e^0+e^5=\sqrt2\, (dv+{1\over2} h du)$ and $e^i=dy^i$ after a relabelling of the transverse coordinates $y$ of the spacetime. For example
$k^{rs}=\langle\eta^r,\eta^s\rangle_H\, (e^0-e^5)$ and so on.

It remains to specify the $k$-form bilinears $\phi^{rs}$.  It turns out that these span all the constant  forms  on the transverse space of the pp-wave up and including those of degree 4.  To see this  decompose $\phi^{rs}=e^\natural\wedge \alpha^{rs}+ \beta^{rs}$, where $\alpha^{rs}$ and $\beta^{rs}$ have components only along the directions transverse to $e^\natural$.   The tensor product of two Majorana $\mathfrak{spin}(8)$ representations, $\Delta_{\bf 16}$, can be decomposed as
\bea
\Delta_{\bf 16}\otimes \Delta_{\bf 16}=\oplus_{k=0}^8 \Lambda^k(\bR^8)~.
\eea
Therefore the forms $\beta^{rs}$ which are up to degree 4  span all forms of the same degree on $\bR^8$ subspace transverse to $e^\natural$. On the other hand the Hodge duals of the forms $\alpha^{rs}$ span all forms of degree 5 and higher in $\bR^8$. Thus  the space of all bilinears of a pp-wave spans a $2^8$-dimensional vector space.

As for the pp-waves we have been considering the 4-form field strength $F$ vanishes, all the form bilinears are covariantly constant with respect to the Levi-Civita connection. As a result all of them generate symmetries for the spinning particle probe action (\ref{sgact}).  The associated conserved charges are given in (\ref{ccalpha}).  They also generate symmetries for string probes as well similar to those investigated in \cite{lggpjp}. The algebra of symmetries can be of W-type and has been described in \cite{phgp}.



\subsection{Hidden symmetries and the KK-monopole}

Choosing the worldvolume directions of the KK-monopole  along $012567\natural$ and allowing $h$ in (\ref{kk}) to be any multi-centred harmonic function as in (\ref{harm}) with $n=3$,   the Killing spinors $\epsilon=\epsilon_0$ of the background satisfy $\Gamma_{3489}\epsilon_0=\pm\epsilon_0$, where $\epsilon_0$ is a constant spinor. To solve this condition with the plus sign, we shall use spinorial geometry and write $\epsilon_0=\eta^1+e_{34}\wedge \lambda^1+e_{3}\wedge \eta^2+e_4\wedge \lambda^2$, where $\eta$ and $\lambda$ are Dirac spinors of  $\mathfrak{spin}(6,1)$, i.e $\eta, \lambda \in \Lambda^{\ast}(\mathbb{C} \langle e_1,e_2,e_5\rangle).$  To begin, let us assume that $\epsilon_0$ is a complex spinor and  impose the reality condition at the end. Then $\Gamma_{3489}\epsilon_0=\epsilon_0$  implies that $\eta^2=\lambda^2=0$ and so $\epsilon_0=\eta+ e_{34}\wedge \lambda$, where $\eta=\eta^1$ and $\lambda=\lambda^1$.  The reality condition on $\epsilon_0$, $\Gamma_{6789}\ast\epsilon_0=\epsilon_0$, implies that $\lambda=-\Gamma_{67}\eta^{\ast}$. Therefore the spinors that solve the Killing spinor condition are
\bea
\epsilon_0=\eta-e_{34}\wedge \Gamma_{67}\eta^{\ast}~,
\eea
where $\eta$ is any  Dirac $\mathfrak{spin}(6,1)$ spinor.

The non-vanishing Killing spinors bilinears read

\bea
&&f^{rs}=2\mathrm{Re}\langle \eta^{r}, \eta^{s}\rangle~,~~~
k^{rs}=2\mathrm{Re}\langle\eta^{r},\Gamma_{a}\eta^{s}\rangle e^a~,
\cr
&&\omega^{rs}={1\over2}\mathrm{Re}\langle\eta^{r},\Gamma_{ab}\eta^{s}\rangle e^{a}\wedge e^b
-2\mathrm{Re}\langle\eta^{r},\lambda^{s}\rangle (e^{3}\wedge e^4-e^8\wedge e^9)
\cr
&&\qquad
-2\mathrm{Im}\langle\eta^{r},\eta^{s}\rangle (e^{3}\wedge e^8+e^4\wedge e^9)
-2\mathrm{Im}\langle\eta^{r},\lambda^{s}\rangle (e^{3}\wedge e^9-e^4\wedge e^8)~,
\cr
&&\varphi^{rs}={1\over3}\mathrm{Re}\langle\eta^{r},\Gamma_{abc}\eta^{s}\rangle e^a \wedge e^b \wedge e^c
-2\mathrm{Im}\langle\eta^{r},\Gamma_{a}\eta^{s}\rangle(e^3\wedge e^8+e^4 \wedge e^9)\wedge e^a
\cr
&&\qquad
-2\mathrm{Im}\langle\eta^{r},\Gamma_{a}\lambda^{s}\rangle(e^3\wedge e^9-e^4\wedge e^8)\wedge e^a~,
\cr
&&\theta^{rs}={1\over12}\mathrm{Re}\langle\eta^{r},\Gamma_{abcd}\eta^{s}\rangle e^a\wedge e^{b}\wedge e^c \wedge e^d
-\mathrm{Re}\langle\eta^{r},\Gamma_{ab}\lambda^{s}\rangle e^a\wedge e^b \wedge (e^3 \wedge e^4-e^8 \wedge e^9)
\cr
&&\qquad
-\mathrm{Im}\langle\eta^{r},\Gamma_{ab}\eta^{s}\rangle e^a\wedge e^b \wedge (e^3 \wedge e^8+ e^4 \wedge e^9)
-\mathrm{Im}\langle\eta^{r},\Gamma_{ab}\lambda^{s}\rangle e^a\wedge e^b \wedge (e^3 \wedge e^{9}-e^4 \wedge e^8 )
\cr
&&\qquad
+2\mathrm{Re}\langle\eta^{r},\eta^{s}\rangle e^3\wedge e^4 \wedge e^8 \wedge e^9~,
\cr
&&\tau^{rs}={1\over60}\mathrm{Re}\langle\eta^{r},\Gamma_{a_1\dots a_5}\eta^{s}\rangle e^{a_1}\wedge \dots \wedge e^{a_5}
+2\mathrm{Re}\langle\eta^{r},\Gamma_{a}\eta^{s}\rangle e^a\wedge e^3 \wedge e^4 \wedge e^8 \wedge e^9
\cr
&&\qquad
-{1\over3}\mathrm{Re}\langle\eta^{r},\Gamma_{abc}\lambda^{s}\rangle e^a\wedge e^b \wedge e^c \wedge (e^3\wedge e^4-e^8\wedge e^9)
\cr
&&\qquad
-{1\over3}\mathrm{Im}\langle\eta^{r},\Gamma_{abc}\eta^{s}\rangle e^a\wedge e^b \wedge e^c \wedge (e^3\wedge e^8+e^4\wedge e^9)
\cr
&&\qquad
-{1\over3}\mathrm{Im}\langle\eta^{r},\Gamma_{abc}\lambda^{s}\rangle e^a\wedge e^b \wedge e^c \wedge (e^3\wedge e^9-e^4\wedge e^8)~,
\label{kkbi}
\eea
where $\langle\cdot, \cdot\rangle$ is the Dirac inner product,  $a, b, c=0, 1, 2, 5, 6, 7,\natural$, $e^a=dx^a$, and $e^i$, $i=3,4,8,9$, is an orthonormal frame of  $g_{(4)}$ in (\ref{kk}), e.g.
\bea
e^3=h^{-{1\over2}} (d\rho+\omega)~,~~~e^4=h^{{1\over2}} dy^4~,~~~e^7=h^{{1\over2}} dy^7~,~~~e^8=h^{{1\over2}} dy^8~,
\eea
after a relabelling of the coordinates of the spacetime.
 The bilinears
of the spinors $\eta$ span all real forms on the worldvolume $\bR^{6,1}$ of the KK-monopole solution. The argument is similar to that produced for the pp-wave.

As for the KK-monopole solution the 4-form field strength vanishes $F=0$, a consequence of the TCFH is that all the form bilinears  in (\ref{kkbi}) are covariantly constant with respect
to the Levi-Civita connection.  As a result they generate symmetries (\ref{svar}) for the spinning particle probe (\ref{sgact}).  The conserved charges
are given in (\ref{ccalpha}).  The algebra of symmetries can be a W-type of algebra \cite{phgp}.

\subsection{Hidden symmetries and  the M2-brane}

Choosing the M2-brane worldvolume directions along $05\natural$, the Killing spinors of the solution are $\epsilon= h^{-{1\over6}} \epsilon_0$, where $\epsilon_0$ is a constant spinor satisfying the condition $\Gamma_{05\natural}\epsilon_0=\pm\epsilon_0$ and $h$ is a (multi-centred) harmonic function as in (\ref{harm}) with $n=8$. To solve the condition with the plus sign use spinorial geometry to write
$\epsilon_0=\eta+e_5\wedge \lambda$, where $\eta, \lambda\in \Lambda^*(\bR\langle e_1, e_2, e_3, e_4\rangle)$.  Then the condition  $\Gamma_{05\natural}\epsilon_0=\epsilon_0$ implies that $\eta, \lambda\in \Lambda^{\mathrm{ev}}(\bR\langle e_1, e_2, e_3, e_4\rangle)$, i.e. $\eta, \lambda$ are Majorana-Weyl $\mathfrak{spin}(8)$ spinors, where the reality condition is imposed with the anti-linear map $\Gamma_{6789}*$.

 Using the solution of the condition on the Killing spinors and setting  $\phi^{rs}=h^{-{1\over3}} \mathring \phi^{rs}$ for for all bilinears $\phi^{rs}$, one can easily  find
\bea
&& \mathring f^{rs}
=-\langle\eta^r,\lambda^s\rangle_H+\langle\lambda^r,\eta^s\rangle_H~,~~~
\cr
&&\mathring k^{rs}=(\langle\eta^r,\eta^s\rangle_H+\langle\lambda^r,\lambda^s\rangle_H)e^0
+(-\langle\eta^r,\eta^s\rangle_H+\langle\lambda^r,\lambda^s\rangle_H)e^5
+(\langle\eta^r,\lambda^s\rangle_H+\langle\lambda^r,\eta^s\rangle_H)e^\natural~,
\cr
&&\mathring\omega^{rs}=(\langle\eta^r,\lambda^s\rangle_H+\langle\lambda^r,\eta^s\rangle_H)e^{0}\wedge e^5
+(\langle\eta^r,\eta^s\rangle_H-\langle\lambda^r,\lambda^s\rangle_H)e^{0}\wedge e^\natural
\cr
&&\qquad
+(-\langle\eta^r,\eta^s\rangle_H-\langle\lambda^r,\lambda^s\rangle_H)e^{5}\wedge e^\natural
+{1\over2}(-\langle\eta^r,\Gamma_{ij}\lambda^s\rangle_H+\langle\lambda^r,\Gamma_{ij}\eta^s\rangle_H)e^{i}\wedge e^j~,
\cr
&&\mathring\varphi^{rs}={1\over2}(\langle\eta^r,\Gamma_{ij}\eta^s\rangle_H+\langle\lambda^r,\Gamma_{ij}\lambda^s\rangle_H)e^{0}\wedge e^i \wedge e^j
+{1\over2}(-\langle\eta^r,\Gamma_{ij}\eta^s\rangle_H+\langle\lambda^r,\Gamma_{ij}\lambda^s\rangle_H)e^{5}\wedge e^i \wedge e^j
\cr
&&\qquad
+{1\over2}(\langle\eta^r,\Gamma_{ij}\lambda^s\rangle_H+\langle\lambda^r,\Gamma_{ij}\eta^s\rangle_H)e^{\natural}\wedge e^i \wedge e^j
+(-\langle\eta^r,\lambda^s\rangle_H+\langle\lambda^r,\eta^s\rangle_H)e^{0}\wedge e^5 \wedge e^\natural~,
\cr
&&\mathring\theta^{rs}={1\over2}\Big((\langle\eta^r,\Gamma_{ij}\eta^s\rangle_H+\langle\lambda^r,\Gamma_{ij}\lambda^s\rangle_H)e^{0}\wedge e^5
+(\langle\eta^r,\Gamma_{ij}\eta^s\rangle_H-\langle\lambda^r,\Gamma_{ij}\lambda^s\rangle_H)e^{0}\wedge e^\natural
\cr
&&\qquad
-(\langle\eta^r,\Gamma_{ij}\eta^s\rangle_H+\langle\lambda^r,\Gamma_{ij}\lambda^s\rangle_H)e^{5}\wedge e^\natural\Big) \wedge e^i \wedge e^j
\cr
&&\qquad
-{1\over4!}(\langle\eta^r,\Gamma_{i_1\dots i_4}\lambda^s\rangle_H-\langle\lambda^r,\Gamma_{i_1\dots i_4}\eta^s\rangle_H)e^{i_1}\wedge \dots \wedge e^{i_4}~,
\cr
&&\mathring\tau^{rs}={1\over2}(-\langle\eta^r,\Gamma_{ij}\lambda^s\rangle_H+\langle\lambda^r,\Gamma_{ij}\eta^s\rangle_H)e^{0}\wedge e^5 \wedge e^\natural \wedge e^i \wedge e^j
+{1\over4!}\Big(\langle\eta^r,\Gamma_{i_1\dots i_4}\eta^s\rangle_H (e^{0}-e^5)
\cr
&&\qquad
+\langle\lambda^r,\Gamma_{i_1\dots i_4}\lambda^s\rangle_H (e^0+ e^5)
+(\langle\eta^r,\Gamma_{i_1\dots i_4}\lambda^s\rangle_H+\langle\lambda^r,\Gamma_{i_1\dots i_4}\eta^s\rangle_H) e^{\natural}\Big)\wedge e^{i_1}\wedge \dots \wedge e^{i_4}~,
\eea
where $(e^a, e^i)$ is the pseudo-orthonormal frame  with $e^a=h^{-1/3} d\sigma^a$,  $a=0,5,\natural$, and $e^i=h^{1/6} dy^i$  $i,j,k,\ell=1, 2, 3, 4, 6, 7, 8, 9$, after an appropriate relabelling of the coordinates of the spacetime.  As the product of two positive chirality Majorana-Weyl $\mathfrak{spin}(8)$ representations, $\Delta^+_{\bf 8}$, is decomposed as
\bea
\otimes^2 \Delta^+_{\bf 8}=\Lambda^0(\bR^8)\oplus \Lambda^2(\bR^8)\oplus \Lambda^{4+}(\bR^8)~,
\label{dec8}
\eea
it is expected that the form bilinears above span all the 0-, 2- and self-dual 4-forms along the transverse directions of the M2-brane.

It remains to find which of the above form bilinears  are KY tensors with respect to the Levi-Civita connection so that generate symmetries for the spinning particle probe (\ref{sgact}). To begin
as the 1-form bilinears $k$ are   Killing they generate symmetries for the action   (\ref{sgact}) and the associated  conserved charges are given  in
(\ref{ccalpha}).  For the  bilinear $\omega$ to be a KY form, it is required that the terms in the TCFH connection that are proportional to $F$ as well as those that in the TCFH that contain explicitly the spacetime metric $g$ must vanish. After some investigation, these terms vanish provided that the components, $\tau_{abcij}$, of the form bilinear $\tau$ are zero, $\tau_{abcij}=0$.  This in turn  implies that $\omega_{ij}=0$.  Setting $\omega_{ij}=0$,   $\omega={1\over2} \omega_{ab} e^a\wedge e^b$ is a KY tensor and  generates a symmetry transformation (\ref{svar}) for the action (\ref{sgact}) with  associated conserved charge given in (\ref{ccalpha}).  Note that
$\omega$ has components only along the  worldvolume directions of the M2-brane.  There are Killing spinors such that $\omega\not=0$, even though $\omega_{ij}=0$, as a consequence of the
decomposition  (\ref{dec8}). A similar investigation reveals that  $\tau$ cannot be a KY form as the conditions arising from the analysis of the TCFH imply
that $\tau=0$.

Next $\varphi$ is a KY form with respect to the Levi-Civita connection provided that the terms proportional to $F$ in the TCFH connection as well as those in the TCFH
that contain explicitly the spacetime metric $g$ vanish. This is the case provided that the components, $\theta_{abij}$, of $\theta$ vanish, $\theta_{abij}=0$.
This in turn implies that $\varphi_{aij}=0$.  Therefore $\varphi={1\over6} \varphi_{abc} e^a\wedge e^b\wedge e^c$ is a KY form and so generates a symmetry for the spinning particle probe action  (\ref{sgact}) with conserved charge (\ref{ccalpha}). Note again that the KY form $\varphi$ has components only along the worldvolume directions of M2-brane and that there are Killing spinors such that $\varphi\not=0$ even though $\varphi_{aij}=0$ as a consequence of (\ref{dec8}).
A similar investigation concludes that $\theta$, as $\tau$,  cannot be a KY form.

\subsection{Hidden symmetries and the M5-brane}

Choosing the worldvolume directions of the M5-brane along $012567$, the Killing spinors of the background are $\epsilon= h^{-{1\over12}} \epsilon_0$, where the constant spinor $\epsilon_0$ satisfies the condition $\Gamma_{3489\natural}\epsilon_0=\pm\epsilon_0$ and $h$ is a multi-centred harmonic function as in (\ref{harm}) with $n=5$. To continue it is convenient to solve the condition on $\epsilon_0$ with a plus sign by taking $\epsilon_0$ to be complex and impose the reality condition on $\epsilon_0$ at the end. Indeed for $\epsilon_0$  complex, one can  use spinorial geometry to write $\epsilon_0=\eta^1+e_{34}\wedge \lambda^1+e_{3}\wedge \eta^2+e_4\wedge \lambda^2$, where $\eta^1, \eta^2, \lambda^1, \lambda^2\in \Lambda^*(\bC\langle e_1,e_2,e_5\rangle)$.  Then the condition $\Gamma_{3489\natural}\epsilon_0=\epsilon_0$ implies that $\eta^1, \eta^2, \lambda^1, \lambda^2\in \Lambda^{\mathrm{ev}}(\bC\langle e_1,e_2,e_5\rangle)$, i.e.
$\eta^1, \eta^2, \lambda^1, \lambda^2$ are positive chirality  spinors of $\mathfrak{spin}(5,1)$. Next imposing the reality condition on $\epsilon_0$,
$\Gamma_{6789}\ast\epsilon_0=\epsilon_0$, one finds that $\lambda^1=-\Gamma_{67}(\eta^{1})^{\ast}$ and $\lambda^2=-\Gamma_{67}(\eta^{2})^{\ast}$.
Hence the spinors that solve the Killing spinor condition are
\bea
\epsilon_0=\eta^1-e_{34}\wedge \Gamma_{67}(\eta^{1})^{\ast}+e_3\wedge \eta^2-e_4\wedge \Gamma_{67}(\eta^{2})^{\ast}~,
\label{m5ks}
\eea
where $\eta^1, \eta^2$ are any positive chirality $\mathfrak{spin}(5,1)$ spinors. The form bilinears of the M5-brane expressed in terms of the $\eta^1$ and $\eta^2$ spinors can be found in appendix \ref{apa}.

The 1-form bilinears $k^{rs}$ are isometries and so generate symmetries for the spinning particle probe action (\ref{sgact}). Next for the bilinear $\omega$ to be a KY tensor, and so generate a symmetry for the spinning particle probe (\ref{sgact}), the term that contains  $F$ in the minimal TCFH connection  ${\cal D}^{\cal F}$ and the term proportional to the spacetime metric $g$ in the TCFH (\ref{11dtcfh}) must vanish. This is the case provided that the component, $\tau_{ijk\ell a}$, of $\tau$ vanishes. However this in turn implies that $\omega=0$ and so $\omega$ does not generate a symmetry.   It turns out  $\theta$, like $\omega$, does not generate a symmetry for the probe (\ref{sgact}) because the conditions required by the TCFH for $\theta$ to be a KY form are too restrictive and yield   $\theta=0$.  Next $\varphi$ is a KY form as a consequence of TCFH  provided that $\theta_{iabc}=\theta_{aijk}=0$.  This implies $\varphi_{aij}=0$ and  leaves the possibility that the remaining component of $\varphi$  $\varphi={1\over3!} \varphi_{abc} e^a\wedge e^b\wedge e^c$ is a KY form.  However after some computation one can verify that there are no Killing spinors such that $\varphi\not=0$. A similar conclusion holds for the $\tau$ form bilinear.





\section{Concluding Remarks}

We have presented the TCFH of 11-dimensional supergravity and we have demonstrated that the form bilinears of supersymmetric backgrounds of the theory satisfy a generalisation of the CKY equation with respect to a connection that depends on the 4-form field strength. We have also given the reduced holonomy of the minimal and maximal TCFH connections for generic backgrounds.

As KY forms with respect to the Levi-Civita connection generate symmetries for spinning particle actions, we investigated the question on whether the form bilinears of 11-dimensional supergravity generate symmetries for suitable particle probes propagating on supersymmetric backgrounds. For this we focused on M-branes which include the pp-wave, M2- and M5-brane, and KK-monopole solutions. As all the form bilinears of pp-wave and KK-monopole solutions are covariantly constant with respect to the Levi-Civita connection, they generate symmetries for the spinning particle action with only a metric coupling. For the M2-brane, there are Killing spinors such that the 1-form, 2-form and 3-form bilinears are KY tensors and therefore generate symmetries for the same spinning particle action. For the M5-brane only
the 1-form bilinears generate symmetries for the spinning particle action.

We also took the opportunity to demonstrate the complete integrability of the geodesic flow of spherically symmetric pp-wave, M2- and M5-brane, and KK-monopole solutions.  For this we presented a large class of KS and KY tensors on all these backgrounds. Relativistic particles on these solutions admit an infinite number of symmetries generated by KS tensors. We have also explicitly  given all independent and in involution conserved charges of the geodesic flow on these backgrounds.

Comparing the symmetries required for the integrability of the geodesic flow on M-brane backgrounds with the symmetries generated by the form bilinears, one arrives at the conclusion that these contribute at different sectors in the probe dynamics. If a form bilinear generates a symmetry for a particle probe, it will generate a symmetry  on  all M-brane backgrounds including those that depend on multi-centred harmonic functions. One does not expect that the geodesic flow on generic such backgrounds to be completely integrable.  Therefore generically form bilinears of supersymmetric backgrounds are not responsible for the integrability properties of a probe. Nevertheless generate additional symmetries for probes, e.g. additional worldvolume supersymmetries, which characterise the dynamics.

As the TCFH connection depends on the 4-form field strength of 11-dimensional supergravity, one should also consider spinning particle probes that exhibit a 4-form coupling. The expectation would be that in this way one can better match the TCFH with the conditions for invariance of the probe action under transformations generated by the form bilinears.  Such a probe action has been presented in appendix B.  However under some reasonable assumptions on the couplings and on the transformations constructed from the form bilinears, one finds that the conditions for invariance of the probe action are too strong for M-brane backgrounds and they do not match with those of TCFH.  This does not  exhausts all possibilities of matching an 11-dimensional TCFH with the  conditions required for the form bilinear to generate a symmetry for a spinning particle probe constructed using the results of   \cite{colesgp}.  One can choose different TCFHs associated with the same supergravity theory as well as different probe actions. So it remains an open question whether such a matching of conditions can be achieved in 11 dimensions.

\section*{Acknowledgments}

EPB is supported by the CONACYT, the Mexican Council of Science.

\setcounter{section}{0}
\setcounter{subsection}{0}
\setcounter{equation}{0}

\begin{appendices}

\section{M5-brane bilinears} \label{apa}

Using the solution (\ref{m5ks}) of the condition on the Killing spinors and setting $\phi^{rs}=h^{-{1\over6}} \mathring \phi^{rs}$ for all bilinears $\phi^{rs}$, one can easily  find
\bea
&&\mathring  f^{rs}=0~,~~~\mathring  k^{rs}=2(\mathrm{Re}\langle\eta^{1r},\Gamma_{a}\eta^{1s}\rangle
+\mathrm{Re}\langle\eta^{2r},\Gamma_{a}\eta^{2s}\rangle)\,e^a~,
\cr
&&\mathring \omega^{rs}=2(\mathrm{Re}\langle\eta^{1r},\Gamma_{a}\eta^{2s}\rangle+\mathrm{Re}\langle\eta^{2r},\Gamma_{a}\eta^{1s}\rangle)\,e^{a}\wedge e^3
\cr
&&\qquad
+2(\mathrm{Re}\langle\eta^{1r},\Gamma_{a}\lambda^{2s}\rangle -\mathrm{Re}\langle\eta^{2r},\Gamma_{a}\lambda^{1s}\rangle)\,e^{a}\wedge e^4
\cr
&&\qquad
+2(\mathrm{Im}\langle\eta^{1r},\Gamma_{a}\eta^{2s}\rangle -\mathrm{Im}\langle\eta^{2r},\Gamma_{a}\eta^{1s}\rangle)\,e^{a}\wedge e^8
\cr
&&\qquad
+2(\mathrm{Im}\langle\eta^{1r},\Gamma_{a}\lambda^{2s}\rangle -\mathrm{Im}\langle\eta^{2r},\Gamma_{a}\lambda^{1s}\rangle)\,e^{a}\wedge e^9
\cr
&&\qquad
+2(\mathrm{Re}\langle\eta^{1r},\Gamma_{a}\eta^{1s}\rangle-\mathrm{Re}\langle\eta^{2r},\Gamma_{a}\eta^{2s}\rangle)\, e^a\wedge e^\natural~,
\cr
&&\mathring \varphi^{rs}={1\over3}(\mathrm{Re}\langle\eta^{1r},\Gamma_{abc}\eta^{1s}\rangle+\mathrm{Re}\langle\eta^{2r},\Gamma_{abc}\eta^{2s}\rangle)\, e^a \wedge e^b \wedge e^c
\cr
&&\qquad
-2\mathrm{Re}\langle\eta^{1r},\Gamma_{a}\lambda^{1s}\rangle(e^3\wedge e^4- e^8\wedge e^9)\wedge e^a
\cr
&&\qquad
+2\mathrm{Re}\langle\eta^{2r},\Gamma_{a}\lambda^{2s}\rangle(e^3\wedge e^4+e^8\wedge e^9)\wedge e^{a}
\cr
&&\qquad
-2\mathrm{Im}\langle\eta^{1r},\Gamma_{a}\eta^{1s}\rangle(e^3\wedge e^8+e^4 \wedge e^9)\wedge e^a
\cr
&& \qquad
+2\mathrm{Im}\langle\eta^{2r},\Gamma_{a}\eta^{2s}\rangle(e^3\wedge e^8-e^4\wedge e^9)\wedge e^{a}
\cr
&&\qquad
-2\mathrm{Im}\langle\eta^{1r},\Gamma_{a}\lambda^{1s}\rangle(e^3\wedge e^9-e^4\wedge e^8)\wedge e^a
\cr
&& \qquad
+2\mathrm{Im}\langle\eta^{2r},\Gamma_{a}\lambda^{2s}\rangle(e^3\wedge e^9+e^4\wedge e^8)\wedge e^{a}
\cr
&&\qquad
-2(\mathrm{Re}\langle\eta^{1r},\Gamma_{a}\eta^{2s}\rangle-\mathrm{Re}\langle\eta^{2r},\Gamma_{a}\eta^{1s}\rangle)e^3\wedge e^{\natural}\wedge e^a
\cr
&&\qquad
-2(\mathrm{Re}\langle\eta^{1r},\Gamma_{a}\lambda^{2s}\rangle+\mathrm{Re}\langle\eta^{2r},\Gamma_{a}\lambda^{1s}\rangle)e^4\wedge e^{\natural}\wedge e^a
\cr
&&\qquad
-2(\mathrm{Im}\langle\eta^{1r},\Gamma_{a}\eta^{2s}\rangle+\mathrm{Im}\langle\eta^{2r},\Gamma_{a}\eta^{1s}\rangle)e^8\wedge e^{\natural}\wedge e^a
\cr
&&\qquad
-2(\mathrm{Im}\langle\eta^{1r},\Gamma_{a}\lambda^{2s}\rangle+\mathrm{Im}\langle\eta^{2r},\Gamma_{a}\lambda^{1s}\rangle)e^9\wedge e^{\natural} \wedge e^a~,
\cr
&&\mathring \theta^{rs}={1\over3}(\mathrm{Re}\langle\eta^{1r},\Gamma_{abc}\eta^{2s}\rangle+\mathrm{Re}\langle\eta^{2r},\Gamma_{abc}\eta^{1s}\rangle)e^a\wedge e^{b}\wedge e^c \wedge e^3
\cr
&&\qquad
+{1\over3}(\mathrm{Re}\langle\eta^{1r},\Gamma_{abc}\lambda^{2s}\rangle-\mathrm{Re}\langle\eta^{2r},\Gamma_{abc}\lambda^{1s}\rangle)e^a\wedge e^b \wedge e^c \wedge e^{4}
\cr
&&\qquad
+{1\over3}(\mathrm{Im}\langle\eta^{1r},\Gamma_{abc}\eta^{2s}\rangle-\mathrm{Im}\langle\eta^{2r},\Gamma_{abc}\eta^{1s}\rangle)e^a\wedge e^b \wedge e^c \wedge e^{8}
\cr
&&\qquad
+{1\over3}(\mathrm{Im}\langle\eta^{1r},\Gamma_{abc}\lambda^{2s}\rangle-\mathrm{Im}\langle\eta^{2r},\Gamma_{abc}\lambda^{1s}\rangle)e^a\wedge e^b \wedge e^c \wedge e^{9}
\cr
&&\qquad
+{1\over3}(\mathrm{Re}\langle\eta^{1r},\Gamma_{abc}\eta^{1s}\rangle-\mathrm{Re}\langle\eta^{2r},\Gamma_{abc}\eta^{2s}\rangle)e^a\wedge e^b \wedge e^c \wedge e^{\natural}
\cr
&&\qquad
+2(\mathrm{Im}\langle\eta^{1r},\Gamma_{a}\lambda^{2s}\rangle+\mathrm{Im}\langle\eta^{2r},\Gamma_{a}\lambda^{1s}\rangle)e^a\wedge e^3 \wedge e^4 \wedge e^{8}
\cr
&&\qquad
-2(\mathrm{Im}\langle\eta^{1r},\Gamma_{a}\eta^{2s}\rangle+\mathrm{Im}\langle\eta^{2r},\Gamma_{a}\eta^{1s}\rangle)e^a\wedge e^3 \wedge e^4 \wedge e^{9}
\cr
&&\qquad
+2(\mathrm{Re}\langle\eta^{1r},\Gamma_{a}\lambda^{2s}\rangle+\mathrm{Re}\langle\eta^{2r},\Gamma_{a}\lambda^{1s}\rangle)e^a\wedge e^3 \wedge e^8 \wedge e^{9}
\cr
&&\qquad
-2(\mathrm{Re}\langle\eta^{1r},\Gamma_{a}\eta^{2s}\rangle-\mathrm{Re}\langle\eta^{2r},\Gamma_{a}\eta^{1s}\rangle)e^a\wedge e^4 \wedge e^8 \wedge e^{9}
\cr
&&\qquad
-2\mathrm{Re}\langle\eta^{1r},\Gamma_{a}\lambda^{1s}\rangle\,e^a\wedge (e^3 \wedge e^4 \wedge e^{\natural}-e^8 \wedge e^9 \wedge e^{\natural})
\cr
&&\qquad
-2\mathrm{Re}\langle\eta^{2r},\Gamma_{a}\lambda^{2s}\rangle\,e^a\wedge (e^3 \wedge e^4 \wedge e^{\natural}+e^8 \wedge e^9 \wedge e^{\natural})
\cr
&&\qquad
-2\mathrm{Im}\langle\eta^{1r},\Gamma_{a}\eta^{1s}\rangle\,e^a\wedge(e^3 \wedge e^8 \wedge e^{\natural}+ e^4 \wedge e^9 \wedge e^{\natural})
\cr
&&\qquad
-2\mathrm{Im}\langle\eta^{2r},\Gamma_{a}\eta^{2s}\rangle\,e^a\wedge (e^3 \wedge e^8 \wedge e^{\natural}- e^4 \wedge e^9 \wedge e^{\natural})
\cr
&&\qquad
-2\mathrm{Im}\langle\eta^{1r},\Gamma_{a}\lambda^{1s}\rangle\,e^a\wedge(e^3 \wedge e^9 \wedge e^{\natural}-e^4 \wedge e^8 \wedge e^{\natural})
\cr
&&\qquad
-2\mathrm{Im}\langle\eta^{2r},\Gamma_{a}\lambda^{2s}\rangle\,e^a\wedge (e^3 \wedge e^9 \wedge e^{\natural}+ e^4 \wedge e^8 \wedge e^{\natural})~,
\cr
&&
\mathring \tau^{rs}
=-{1\over3}\mathrm{Re}\langle\eta^{1r},\Gamma_{abc}\lambda^{1s}\rangle e^a\wedge e^b \wedge e^c \wedge( e^3\wedge e^4- e^8\wedge e^9)
\cr
&&\qquad
+{1\over3}\mathrm{Re}\langle\eta^{2r},\Gamma_{abc}\lambda^{2s}\rangle\,e^a\wedge e^b \wedge e^c \wedge (e^3\wedge e^4+ e^8\wedge e^9)
\cr
&&\qquad
-{1\over3}\mathrm{Im}\langle\eta^{1r},\Gamma_{abc}\eta^{1s}\rangle e^a\wedge e^b \wedge e^c \wedge (e^3\wedge e^8+e^4\wedge e^9)
\cr
&&\qquad
+{1\over3}\mathrm{Im}\langle\eta^{2r},\Gamma_{abc}\eta^{2s}\rangle\,e^a\wedge e^b \wedge e^c \wedge (e^3\wedge e^8-e^4\wedge e^9)
\cr
&&\qquad
-{1\over3}\mathrm{Im}\langle\eta^{1r},\Gamma_{abc}\lambda^{1s}\rangle e^a\wedge e^b \wedge e^c \wedge (e^3\wedge e^9-e^4\wedge e^8)
\cr
&&\qquad
+{1\over3}\mathrm{Im}\langle\eta^{2r},\Gamma_{abc}\lambda^{2s}\rangle\,e^a\wedge e^b \wedge e^c \wedge (e^3\wedge e^9+e^4\wedge e^8)
\cr
&&\qquad
-{1\over3}(\mathrm{Re}\langle\eta^{1r},\Gamma_{abc}\eta^{2s}\rangle-\mathrm{Re}\langle\eta^{2r},\Gamma_{abc}\eta^{1s}\rangle)e^a\wedge e^b \wedge e^c \wedge e^3\wedge e^\natural
\cr
&&\qquad
-{1\over3}(\mathrm{Re}\langle\eta^{1r},\Gamma_{abc}\lambda^{2s}\rangle+\mathrm{Re}\langle\eta^{2r},\Gamma_{abc}\lambda^{1s}\rangle)e^a\wedge e^b \wedge e^c \wedge e^4\wedge e^\natural
\cr
&&\qquad
-{1\over3}(\mathrm{Im}\langle\eta^{1r},\Gamma_{abc}\eta^{2s}\rangle+\mathrm{Im}\langle\eta^{2r},\Gamma_{abc}\eta^{1s}\rangle)e^a\wedge e^b \wedge e^c \wedge e^8\wedge e^\natural
\cr
&&\qquad
-{1\over3}(\mathrm{Im}\langle\eta^{1r},\Gamma_{abc}\lambda^{2s}\rangle+\mathrm{Im}\langle\eta^{2r},\Gamma_{abc}\lambda^{1s}\rangle)e^a\wedge e^b \wedge e^c \wedge e^9\wedge e^\natural
\cr
&&\qquad
+{2\over5!} \big(\mathrm{Re}\langle \eta^{1r}, \Gamma_{a_1\dots a_5}\eta^{1s}\rangle+\mathrm{Re}\langle \eta^{2r}, \Gamma_{a_1\dots a_5}\eta^{2s}\rangle\big)~e^{a_1}\wedge\dots\wedge e^{a_5}
\cr
&&\qquad
+2(\mathrm{Re}\langle\eta^{1r},\Gamma_{a}\eta^{1s}\rangle-\mathrm{Re}\langle\eta^{2r},\Gamma_{a}\eta^{2s}\rangle)e^a\wedge e^3 \wedge e^4 \wedge e^8\wedge e^9
\cr
&&\qquad
-2(\mathrm{Im}\langle\eta^{1r},\Gamma_{a}\lambda^{2s}\rangle-\mathrm{Im}\langle\eta^{2r},\Gamma_{a}\lambda^{1s}\rangle)e^a\wedge e^3 \wedge e^4 \wedge e^8\wedge e^\natural
\cr
&&\qquad
+2(\mathrm{Im}\langle\eta^{1r},\Gamma_{a}\eta^{2s}\rangle-\mathrm{Im}\langle\eta^{2r},\Gamma_{a}\eta^{1s}\rangle)e^a\wedge e^3 \wedge e^4 \wedge e^9\wedge e^\natural
\cr
&&\qquad
-2(\mathrm{Re}\langle\eta^{1r},\Gamma_{a}\lambda^{2s}\rangle-\mathrm{Re}\langle\eta^{2r},\Gamma_{a}\lambda^{1s}\rangle)e^a\wedge e^3 \wedge e^8 \wedge e^9\wedge e^\natural
\cr
&&\qquad
+2(\mathrm{Re}\langle\eta^{1r},\Gamma_{a}\eta^{2s}\rangle+\mathrm{Re}\langle\eta^{2r},\Gamma_{a}\eta^{1s}\rangle)e^a\wedge e^4 \wedge e^8 \wedge e^9\wedge e^\natural~,
\eea
where after a relabelling of the spacetime coordinates  $e^a= h^{-1/6} d\sigma^a$, $a=0, 1, 2, 5, 6, 7$, and  $e^i= h^{1/3} dy^i$, $i=3,4,8,9,\natural$, is a pseudo-orthonormal frame of the M5-brane metric (\ref{m5}).

\section{Spinning particle probes with 4-form couplings}

Following the use of spinning particle probes on 4- and 5-dimensional minimal supergravity backgrounds that exhibit 2-form couplings \cite{ebgp}, one
may be tempted to generalise these to spinning particle probes that exhibit 4-form couplings.  Such a generalisation is desirable as the TCFH connection
of 11-dimensional supergravity exhibits terms that depend of the 4-form field strength $F$. One way to generalise (\ref{sgact}) is to adapt the general construction
of \cite{colesgp} and introduce a fermonic superfield $\psi$ of mass dimension $[1/2]$.  Insisting for the couplings of the action to be dimensionless, a minimal choice for an action with a 4-form coupling is
\bea
\label{a1}
S=-{1\over 2}\int dt d\theta\,[ig_{\mu\nu}Dx^\mu \partial_tx^\nu-{i\over 12}F_{\mu\nu\rho\sigma}\psi^{\mu\nu\rho}\partial_t x^\sigma+ \beta \psi_{\mu\nu\rho} \nabla \psi^{\mu\nu\rho}]~,
\eea
with $\beta$ a constant which will be specified later,
\bea
\nabla\psi^{\mu\nu\rho}=D\psi^{\mu\nu\rho}+3 Dx^\lambda \Gamma^{[\mu}_{\lambda \mu'}  \psi^{|\mu'|\nu\rho]}~,
\eea
and $\Gamma$ are the Christoffel symbols of the spacetime metric $g$. The numerical coefficient of the coupling $F\psi\partial_tx$ could be arbitrary but the above choice will suffice. Also one could add additional terms in the action like $F\nabla\psi Dx$ which we shall explore  later. Other terms  include couplings of the type
$\nabla F \psi Dx$.  After a superspace partial integration these can be re-expressed in terms of the $F\psi\partial_tx$ and $ F \nabla\psi Dx$ couplings.

The variation of the action (\ref{a1}) can be expressed as
\bea
\delta S=-\int dtd\theta\, [ g_{\mu\nu} \delta x^\mu {\cal S}^\nu+ \Delta \psi_{\mu\nu\rho} {\cal S}^{\mu\nu\rho}]~,
\eea
where
\bea
\Delta\psi^{\mu\nu\rho}=\delta\psi^{\mu\nu\rho}+ 3 \delta x^\lambda \Gamma^{[\mu}_{\lambda \mu'}  \psi^{|\mu'|\nu\rho]}~,
\eea
$\delta x$ and $\delta \psi$ are arbitrary variations of the fields
and
\bea
&&{\cal S}^\mu=-i\nabla_tD x^\mu-{i\over24} \nabla^\mu F_{\nu\rho\sigma\lambda} \psi^{\nu\rho\sigma} \partial_t x^\lambda-{i\over24}\nabla_\lambda F^\mu{}_{\nu\rho\sigma} \psi^{\nu\rho\sigma} \partial_t x^\lambda
\cr
&&\qquad
-{i\over24} F^\mu{}_{\nu\rho\sigma} \nabla_t \psi^{\nu\rho\sigma}+{3\over2} \beta
\psi_{\nu\rho\sigma} D x^\lambda R^\mu{}_{\lambda,}{}^{\nu}{}_{\tau} \psi^{\tau \rho\sigma}~,
\cr
&&
{\cal S}^{\mu\nu\rho}=\beta \nabla \psi^{\mu\nu\rho}-{i\over24} F^{\mu\nu\rho}{}_\lambda \partial_t x^\lambda~,
\label{eom}
\eea
are the equations of motion of $x$ and $\psi$, respectively.

The action (\ref{a1}) is manifestly invariant under one supersymmetry. For a probe described by the action (\ref{a1}) propagating  on an M-brane background with spacetime metric $g$ and 4-form field strength $F$ to exhibit additional symmetries that are generated
by the form bilinears $\omega$ and $\tau$ of 11-dimensional supergravity, one can consider the infinitesimal transformations
\bea
&&\delta x^\mu=\alpha\, \omega^\mu{}_\nu D x^\nu+ \alpha\, c_1\, \omega_{\rho\sigma} \psi^{\mu\rho\sigma}~,
\cr
&&
\delta\psi^{\mu\nu\rho}=\alpha\, \tau^{\mu\nu\rho}{}_{\sigma\lambda} D x^\sigma D x^\lambda+ \alpha\, c_2\,  \tau^{\mu\nu\rho}{}_{\sigma\lambda} \psi^{\sigma\lambda}{}_{\kappa} D x^\kappa~,
\label{infvarx}
\eea
where $\alpha$ is the supersymmetry parameter assigned mass dimension $[-1/2]$ and $c_1$, $c_2$ are constants.  These transformations are the most general ones allowed such that the infinitesimal variations have the same mass dimension as those of the associated fields,  and  $\omega$ and $\tau$ are dimensionless.

For the TCFH on $\omega$ to be interpreted as an invariance condition for the probe action (\ref{a1}), the conditions that arise for the invariance of this action
under the infinitesimal variations (\ref{infvarx}) should match the  TCFH. For this first notice that the equations of motion (\ref{eom}) contain the spacetime curvature $R$. As such terms do not arise in the TCFH,  these  terms in the invariance conditions  must vanish.  This requires that $\beta=0$.  Moreover if the action had contained a $F\nabla\psi Dx$ coupling, this would have given rise to a $F R$ term in the equations of motion. Because the TCFH does not contain such a term, the $F\nabla\psi Dx$ coupling was neglected from the beginning.  The remaining conditions that arise from the invariance of the action (\ref{a1}) with $\beta=0$ under the
infinitesimal transformations (\ref{infvarx}) read
\bea
&&\nabla_\mu \omega_{\nu\rho}-{1\over12} F_{ \mu\lambda\kappa\sigma} \tau^{\lambda\kappa\sigma}{}_{\nu\rho}=0~,~~~c_1\omega_{[\rho\sigma} g_{\nu]\mu}+{1\over24} F_{\rho\sigma\nu \kappa} \omega^\kappa{}_{\mu}=0~,
\cr
&&
(2c_1+c_2) F_{\lambda \kappa_1\kappa_2\kappa_3} \tau^{\kappa_1\kappa_2\kappa_3}{}_{[\rho\sigma} g_{\nu]\mu}+
\nabla_\kappa F_{\nu\rho\sigma\lambda} \omega^\kappa{}_\mu-\nabla_\lambda F_{\nu\rho\sigma \kappa} \omega^\kappa{}_\mu=0~,
\cr
&&
F_{\mu_1\mu_2\mu_3[\nu_1} \omega_{\nu_2\nu_3]}=0~,~~~-\omega_{[\mu_1\mu_2} \nabla_{\mu_3]} F_{\nu_1\nu_2\nu_3\lambda}+\nabla_\lambda F_{\nu_1\nu_2\nu_3[\mu_1} \omega_{\mu_2\mu_3]}=0~.
\label{invconx}
\eea
The first condition matches the expression of the TCFH connection on $\omega$.  However the second condition is rather strong on both M2- and M5-brane backgrounds to admit non-trivial solutions. Moreover this condition persists even if $\beta\not=0$ and the curvature terms are included. This does not exclude the possibility that there may be backgrounds such that the TCFH matches with the conditions (\ref{invconx}) but if this is the case, such examples will be  restricted.

\end{appendices}

\end{document}